\documentclass[abstracton,titlepage,12pt]{scrartcl}
\usepackage{amsmath}
\usepackage{amssymb,latexsym}
\usepackage{graphicx}
\usepackage[usenames]{color}
\usepackage{tcolorbox}
\usepackage[normalem]{ulem}
\usepackage{soul}

\usepackage{float}
 
\floatstyle{boxed}
\newfloat{nicebox}{thp}{lop}
\floatname{nicebox}{Box}

\usepackage{setspace}

\usepackage[left,displaymath,mathlines]{lineno}

\newcommand{\veps}{\varepsilon}

\def\mrm#1{\mathrm{#1}}
\def\fig#1{Fig.\,\ref{#1}}

\def\sec#1{Sec.\,\ref{#1}}
\def\eq#1{Eq.\,(\ref{#1})}

\def\fig#1{Fig.\,\ref{#1}}
\begin{document}


\begin{titlepage}

  \title{Can high risk fungicides be used in mixtures without
    selecting for fungicide resistance?}

\author{Alexey Mikaberidze, \\
 Bruce A. McDonald,\\
Sebastian Bonhoeffer}


\publishers{

\begin{normalsize}

\begin{flushleft}
affiliation: Institute of Integrative Biology, ETH Zurich\\
\vspace{0.5cm}
keywords: epidemiology, plant disease, mathematical model, host-pathogen
interaction, fungicide resistance, population dynamics
\vspace{0.2cm}

\end{flushleft}

\vspace{0.2cm}

corresponding author:\\
Alexey Mikaberidze (alexey.mikaberidze@env.ethz.ch)

\end{normalsize}
}

\date{}

\end{titlepage}

\maketitle

\doublespacing

\begin{abstract}
  Mikaberidze, A., McDonald, B. A., Bonhoeffer, S. 2013. Can high risk
  fungicides be used in mixtures without selecting for fungicide
  resistance? Phytopathology.

  Fungicide mixtures produced by the agrochemical industry often
  contain low-risk fungicides, to which fungal pathogens are fully
  sensitive, together with high-risk fungicides known to be prone to
  fungicide resistance.
  Can these mixtures provide adequate disease control while minimizing
  the risk for the development of resistance?  We present a population
  dynamics model to address this question. We found that the fitness
  cost of resistance is a crucial parameter to determine the outcome
  of competition between the sensitive and resistant pathogen strains
  and to assess the usefulness of a mixture. If fitness costs are
  absent, then the use of the high-risk fungicide in a mixture selects
  for resistance and the fungicide eventually becomes
  nonfunctional. If there is a cost of resistance, then an optimal
  ratio of fungicides in the mixture can be found, at which selection
  for resistance is expected to vanish and the level of disease
  control can be optimized.
\end{abstract}



Fungicide resistance is a prime example of adaptation of a population
to an environmental change, also known as evolutionary rescue
\cite{bego11,hosg11}.
While global climate change is expected to result in a loss of
biodiversity in natural ecosystems, evolutionary rescue is seen as a
mechanism that may mitigate this loss. In the context of crop
protection the point of view is quite the opposite: reducing
adaptation of crop pathogens to chemical disease control would help
stabilize food production. Better understanding of the adaptive
process may help slow or prevent it.
This requires a detailed quantitative understanding of the dynamics of
infection and the factors driving the emergence and development of
fungicide resistance \cite{bogi08}. 
Despite the global importance and
urgency of fungicide resistance, this problem has received relatively
little theoretical consideration (see
\cite{hopa+11a,hagu+07,sh06,pagi+05,sh89,mile+89} and \cite{bogi08}
for a comprehensive review) as compared, for example, to antibiotic
resistance \cite{ozsh+12,boau+01,le01,auan99}. 

In recent years, agrochemical companies have begun marketing mixtures
that contain fungicides with a low-risk of developing resistance with
fungicides that have a high-risk developing of resistance. In extreme
cases the high-risk fungicide is no longer effective against some
common pathogens because resistance has become widespread. For
example, a large proportion of the European population of the
important wheat pathogen \emph{Mycosphaerella graminicola} (recently
renamed \emph{Zymoseptoria tritici}) \cite{orde+11,pask03} is
resistant to strobilurin fungicides \cite{tobr+09}.


A number of previous modeling studies addressed the effect of
fungicide mixtures on selection for fungicide resistance (for example,
\cite{kaje80,sk81,jodo85,sh93,hopa+11a,hopa+13}). Different studies
used different definitions of ``independent action'' (also called
``additivity'' or ``zero interaction'' in the literature) of
fungicides in the mixtures \cite{sh89a} and reported somewhat
different conclusions. One study \cite{sh89a} critically reviewed the
outcomes of these earlier studies and attempted to clarify the
consequences of using different definitions of ``independent
action''. Some studies found that alternations are preferable to
mixtures \cite{kaje80}, while others found that mixtures are
preferable to alternations \cite{sk81}. A more recent study
\cite{hopa+13} addressed this question using a detailed population
dynamics model and found that in all scenarios considered, mixtures to
provided the longest effective life of fungicides as compared to
alternations or concurrent use (when each field receives a single
fungicide, but the fungicides applied differ between the fields). This
study used the Bliss' definition of ``independent action'' of the two
fungicides \cite{bl39} (also called Abbot's formula in the fungicide
literature \cite{ab25}).
%
%


We addressed the question of whether mixtures of low-risk and high
risk fungicides can provide adequate disease control while minimizing
further selection for resistance using a simple population dynamics
model of host-pathogen interaction based on a system of ordinary
differential equations. We found that the fitness cost associated with
resistance mutations is a crucial parameter, which governs the outcome
of the competition between the sensitive and resistant pathogen
strains.
%



A single point mutation associated with fungicide resistance sometimes
makes the pathogen completely insensitive to a fungicide, as is the
case for the G143A mutation giving resistance to strobilurin
fungicides in many fungal pathogens \cite{feto+08,gisi+02}. In many
other cases the resistance is partial, for example, resistance of
\emph{Z. tritici} and other fungi to azole fungicides
\cite{co08,zhst+06}.
Therefore, we considered varying degrees of resistance in our model.


In contrast to our study, resistance in \cite{hopa+11a} was assumed to
bear no fitness costs for the pathogen. It was found that in the
absence of fitness costs the use of fungicide mixtures \emph{delays}
the development of resistance \cite{hopa+11a}. This conclusion is in
agreement with our results (see Appendix\,A.4).
Here we focus on finding conditions under which the selection for the
resistant pathogen strain is \emph{prevented} by using fungicide
mixtures.

\section{Theory and approaches}
\label{sec:model-assump}

\begin{figure}[!ht]
  \centerline{\includegraphics[width=\textwidth]{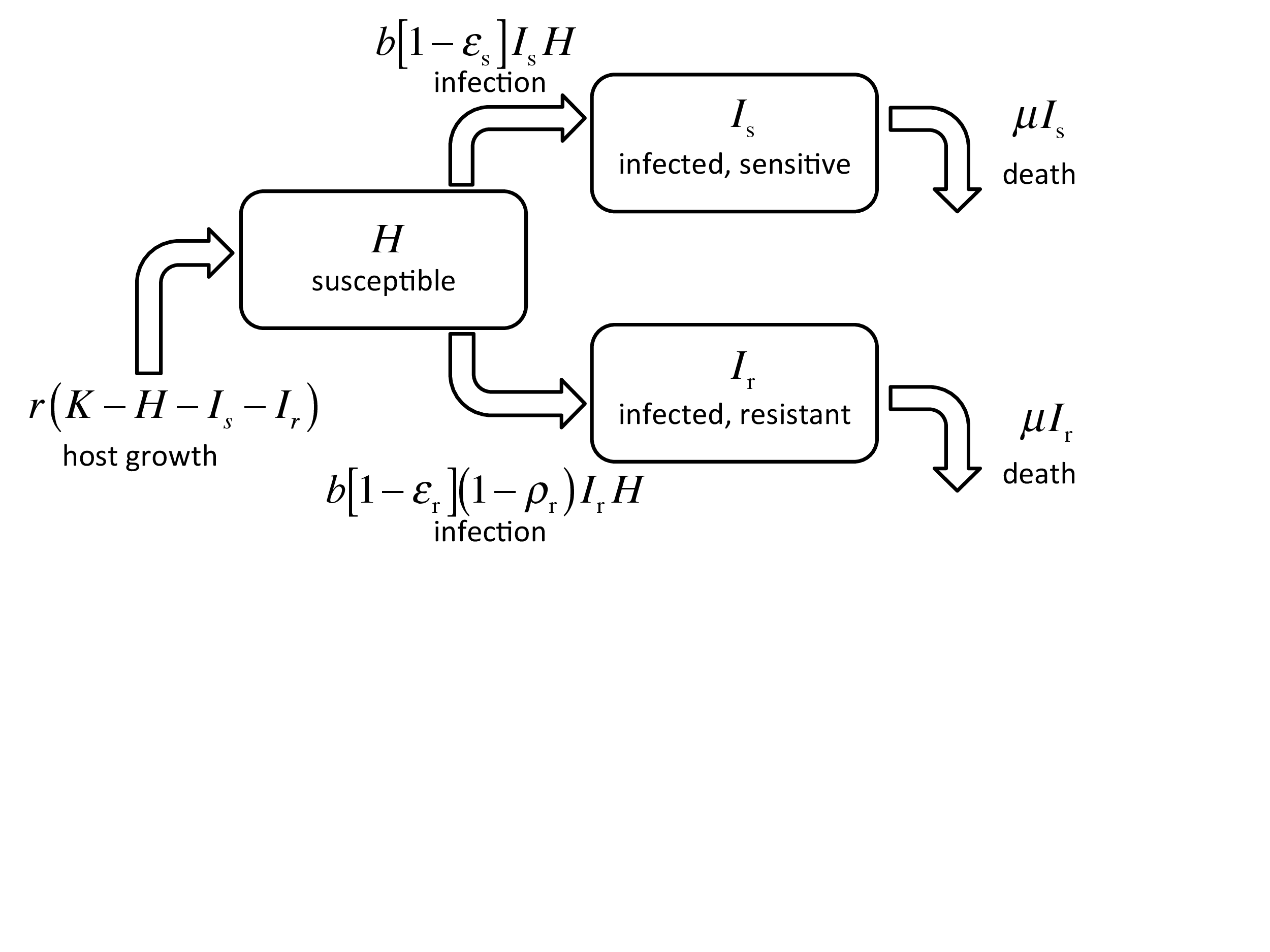}}
 \caption{\doublespacing Scheme of the model in
     Eqs.\,(\ref{eq:1host2fung-gen-1})-(\ref{eq:1host2fung-gen-3}).}
\label{fig:model-scheme}
\end{figure}

We use a deterministic mathematical model of susceptible-infected
dynamics (see \fig{fig:model-scheme})

\begin{align}
\frac{d H}{d t} &= r_H (K - H - I_\mrm{s} - I_\mrm{r} ) -  b \left( \left[ 1 - \veps_\mrm{s}(C,
  r_\mrm{B}) \right] I_\mrm{s} + \left[ 1 - \veps_\mrm{r}(C, r_\mrm{B}) \right] (1-\rho_\mrm{r}) I_\mrm{r} \right) H, \label{eq:1host2fung-gen-1}\\
\frac{d I_\mrm{s}}{d t} & = b \left[ 1 - \veps_\mrm{s}(C,
  r_\mrm{B}) \right] H I_\mrm{s} - \mu  I_\mrm{s}, \label{eq:1host2fung-gen-2}\\ 
\frac{d I_\mrm{r}}{d t} & = b \left[ 1 - \veps_\mrm{r}(C,
  r_\mrm{B}) \right] (1-\rho_\mrm{r}) H I_\mrm{r} - \mu  I_\mrm{r}. \label{eq:1host2fung-gen-3}
\end{align}
The model has three compartments: healthy hosts $H$, hosts
infected by a sensitive pathogen strain $I_\mrm{s}$, hosts infected by
a resistant pathogen strain $I_\mrm{r}$; and is similar to the models
described in \cite{bogi08,hagu+07}. The subscript ``s'' stands for the
sensitive strain and the subscript ``r'' stands for the resistant
strain.
The quantities $H$, $I_\mrm{s}$ and $I_\mrm{r}$, represent the total amount of the
corresponding host tissue within one field, which could be leaves,
stems or grain tissue, depending on the specific host-pathogen interaction.
Healthy hosts $H$ grow with the rate $r_H$. Their growth is
limited by the ``carrying capacity'' $K$, which may imply limited
space or nutrients.
Furthermore, healthy hosts may be infected by the sensitive
pathogen strain and transformed into infected hosts in the compartment
$I_\mrm{s}$ with the transmission rate $b$. This is a compound
parameter given by the product of the sporulation rate of the infected
tissue and the probability that a spore causes new infection.
Healthy hosts may also be infected by the resistant pathogen strain
and transformed into infected hosts in the compartment $I_\mrm{r}$. In
this case, resistant mutants suffer a fitness cost $\rho_\mrm{r}$
which affects their transmission rate such that it becomes equal to $b
(1 - \rho_\mrm{r})$.
The corresponding terms in
Eqs.\,(\ref{eq:1host2fung-gen-1})-(\ref{eq:1host2fung-gen-3}) are
proportional to the amount of the available healthy tissue $H$ and
to the amount of the infected tissue $I_\mrm{s}$ or
$I_\mrm{r}$. Infected host tissue loses its infectivity at a rate
$\mu$, where $\mu^{-1}$ is the characteristic infectious period.

Since our description is deterministic we do not take into account the
emergence of new resistance mutations but assume that the resistant
pathogen strain is already present in the
population. Therefore, when ``selection for resistance'' is discussed
below,
we refer to the process of winning the competition by this existing
resistant strain due to its higher fitness with respect to the
sensitive strain in the presence of fungicide treatment.
Emergence of new resistance mutations is a different problem, which
goes beyond the scope of our study and requires stochastic simulation
methods. We do not consider the possibility of double resistance in
the model, but by preventing selection for single resistance
as described here, one would also diminish the probability of the
emergence of double resistance for both sexually and asexually
reproducing pathogens (see Appendix\,A.7).

We consider two fungicides A and B. Fungicide A is the high-risk
fungicide, to which the resistant pathogen strain exhibits a variable
degree of resistance. However, the sensitive strain is fully sensitive to
fungicide A. Fungicide B is the low-risk fungicide, i.\,e. both pathogen
strains are fully sensitive to it.
We compare the effects of the fungicide A applied alone,
fungicide B applied alone and the effect of their mixture in
different proportions.

We assume that the fungicides will decrease the pathogen transmission
rate $b$ [see the
expression in square brackets in \eq{eq:1host2fung-gen-2}, \eq{eq:1host2fung-gen-3}]. 
For example, application of a fungicide could result in production of
spores that are deficient essential metabolic products such as ergosterol or
$\beta$-tubulin. Consequently, these spores would
likely have a lower success rate in causing new infections.
Spores of sensitive strains of \emph{Z. tritici} produced shorter germ
tubes when exposed to azoles \cite{leal+07}. Spores that produce
shorter germ tubes are less likely to find and penetrate stomata,
hence are less likely to give rise to new infections. Protectant
activity of fungicides will also reduce the transmission rate $b$
\cite{wowi01,pf06}. These studies \cite{wowi01,pf06} also reported
that fungicide application leads to a reduction in the number of
spores produced. This outcome can be attributed to the fungicide
decreasing the sporulation rate and thus affecting $b$ or decreasing
the infectious period and thus affecting $\mu$, or both of these
effects. More detailed measurements are often needed to distinguish
between these different effects.



When only one fungicide applied, the reduction of the transmission rate is described by
\begin{equation}\label{eq:eps-1fungic}
  \veps_\mrm{A}(C_\mrm{A}) = k_\mrm{kA} \frac{C_\mrm{A}}{C_\mrm{A} + C_\mrm{50A}},
\end{equation}
for the fungicide A, and by
\begin{equation}\label{eq:eps-1fungic}
 \veps_\mrm{B}(C_\mrm{B}) = k_\mrm{kB} \frac{C_\mrm{B}}{C_\mrm{B} + C_\mrm{50B}},
\end{equation}
for the fungicide B.
These functions grow with the fungicide doses $C_\mrm{A}$,
$C_\mrm{B}$ and saturate to values $k_\mrm{kA}$, $k_\mrm{kB}$,
respectively, which are the maximum reductions in the transmission
rate (or efficacies). This functional form was used before in the fungicide
resistance literature \cite{hagu+07,gugi99}. We also performed the
analysis for the exponential fungicide action more common in plant
pathology and obtained qualitatively similar results. The reason for
choosing the function in \eq{eq:eps-1fungic} was that it made possible
to obtain all the results analytically.
The parameters $C_\mrm{50A}$, $C_\mrm{50B}$ represent the fungicide dose at
which half of the maximum effect is achieved.
 These parameters can always be made equal by rescaling the
concentration axis for one of the fungicides. Hence, we set $C_\mrm{50A}=C_\mrm{50B}=C_\mrm{50}$.

%
%


We next determine the effect of a mixture of two fungicides
according to the Loewe's definition of additivity (or non-interaction)
\cite{be89} (an equivalent graphic procedure is known as the Wadley
method in the fungicide literature \cite{lebe+86}). It is based on the
notion that a compound cannot interact pharmacologically with
itself. A sham mixture of a compound A with itself can be created and
its effect used as a reference point for assessing of whether the
components of a real mixture interact pharmacologically. When the two
compounds A and B have the same effect as the sham mixture of the
compound A with itself, they are said to have no interaction (or an
additive interaction). In this case, the isobologram equation
\begin{equation}\label{eq:isobol}
C_\mrm{A}/C_\mrm{Ai} + C_\mrm{B}/C_\mrm{Bi} = 1
\end{equation}
holds (see Sec. VA of \cite{be89} for the derivation).
Here, $C_\mrm{A}$ and $C_\mrm{B}$ are the doses of the compounds A and B,
respectively, when applied in the mixture; $C_\mrm{Ai}$ is the
isoeffective dose of the compound A, that is the dose at which
compound A alone has the same effect as the mixture; and $C_\mrm{Bi}$ is the
isoeffective dose of the compound B.
If the mixture of A and B has a larger effect than the
zero-interactive sham mixture, then $C_\mrm{A}/C_\mrm{Ai} +
C_\mrm{B}/C_\mrm{Bi} < 1$ and the two compounds are said to interact
synergistically. On the contrary, when the mixture of A and B has a
smaller effect than the zero-interactive sham mixture,
$C_\mrm{A}/C_\mrm{Ai} + C_\mrm{B}/C_\mrm{Bi} > 1$ and the two
compounds interact antagonistically.

Using the dose-response dependencies of each fungicide when applied
alone, \eq{eq:eps-1fungic} and \eq{eq:isobol}, we derive the dose-response function
for the combined effect of the two fungicides on the sensitive
pathogen strain in the case of no pharmacological interaction (see Sec. VIB of \cite{be89} for the derivation):
\begin{align}\label{eq:veps-sens-addit}
\veps_\mrm{s}(C_\mrm{A}, C_\mrm{B}) =  \frac{ k_\mrm{kA} C_\mrm{A} +
 k_\mrm{kB} C_\mrm{B}} { C_\mrm{A} + C_\mrm{B} + C_{50}}.
\end{align}
Similarly, we determine the combined effect of the two fungicides on the
resistant pathogen strain still without pharmacological
interaction:
\begin{align}\label{eq:veps-res-addit}
\veps_\mrm{r}(C_\mrm{A}, C_\mrm{B}) = \frac{ k_\mrm{kA} \alpha C_\mrm{A} +
k_\mrm{kB}  C_\mrm{B}} { \alpha C_\mrm{A} + C_\mrm{B} + C_{50}},
\end{align}
where we introduced $\alpha$, the degree of sensitivity of the resistant strain 
to the fungicide A (the high-risk fungicide). At $\alpha=0$ the pathogen is fully resistant to
fungicide A and the effect of the mixture $\veps_\mrm{r}(C_\mrm{A},
C_\mrm{B})$ in \eq{eq:veps-res-addit} does not depend on its
dose $C_\mrm{A}$, while at $\alpha = 1$ the pathogen is fully
sensitive to fungicide A.

The expression in \eq{eq:veps-sens-addit} and \eq{eq:veps-res-addit}
are only valid in the range of fungicide concentrations, over which
isoeffective concentrations can be determined for both
fungicides. Here, the isoeffective concentration is the concentration
of a fungicide applied alone that has the same effect as the
mixture. This requirement means that we are only able to consider the
effect of the mixture at a sufficiently low total concentration: $C =
C_\mrm{A} + C_\mrm{B}<k_\mrm{kB} C_{50}/(k_\mrm{kA} - k_\mrm{kB})/(1 -
r_\mrm{B})$.

Next, we introduce deviations from the additive pharmacological
interaction. There are several ways to do this, usually by adding an
interaction term to the isobologram equation \cite{grbr95}. We chose a
specific form of the interaction term, which is proportional to the
square root of the product of the concentrations of the two compounds
[Eq.\,(28) in \cite{grbr95}]. Assuming
$k_\mrm{kA}=k_\mrm{kB}=k_\mrm{k}$, this form allows for a simple
analytical expression for the effect of the combination on the
sensitive strain
\begin{align}\label{eq:veps-mix-comp-sens}
\veps_\mrm{s}(C, r_\mrm{B}) = k_\mrm{k} \frac{ C} { C + C_{50}/ \gamma_\mrm{s}},
\end{align}
and on the resistant strain
\begin{align}\label{eq:veps-mix-comp-res}
\veps_\mrm{r}(C, r_\mrm{B}) = k_\mrm{k} \frac{C} { C  + C_{50}/ \gamma_\mrm{r}}.
\end{align}
Here $C=C_\mrm{A}+C_\mrm{B}$, where $C_\mrm{A}$ is the dose of the fungicide A
and $C_\mrm{B}$ is the dose of the fungicide B, $r_\mrm{B} = C_\mrm{B}/C$
is the proportion of the fungicide B in the mixture and
\begin{equation}\label{eq:gammas}
\gamma_\mrm{s} = 1 + u \sqrt{r_\mrm{B}(1-r_\mrm{B})},
\end{equation}
\begin{equation}\label{eq:gammar}
\gamma_\mrm{r} = \alpha (1 - r_\mrm{B}) + r_\mrm{B}  + u \sqrt{\alpha r_\mrm{B}(1-r_\mrm{B})}
\end{equation}
are the parameters which modify $C_{50}$ due to pharmacological
interaction and partial
resistance. Eqs.\,(\ref{eq:veps-mix-comp-sens}),
(\ref{eq:veps-mix-comp-res}) are obtained from \eq{eq:isobol} with an
interaction term added and the dose-response functions of each
fungicide when applied alone, \eq{eq:eps-1fungic}.
The degree of pharmacological interaction is characterized by the
parameter $u$. At $u=0$ the fungicides do not interact and
Eqs\,(\ref{eq:veps-mix-comp-sens}), (\ref{eq:veps-mix-comp-res}) are
the same as Eqs\,(\ref{eq:veps-sens-addit}),
(\ref{eq:veps-res-addit}). The case when $u>0$ represents synergy: the
interaction term proportional to $u$ in \eq{eq:veps-mix-comp-sens} and
\eq{eq:veps-mix-comp-res} is positive and it reduces the value of
$C_{50}$, meaning that the same effect can be achieved at a lower dose
than at $u=0$. The case when $u<0$ corresponds to antagonism (see
Appendix\,A.1). Note, that the interaction term is proportional to
$\sqrt{r_\mrm{B}(1-r_\mrm{B})}$. This functional form guarantees that
it vanishes, whenever only one of the compounds is used,
i.\,e. $r_\mrm{B}=0$ or $r_\mrm{B}=1$.

%

 In order to make clear the questions we ask and the assumptions we
 make, we consider the dynamics of the frequency of the resistant strain
 $p(t) = I_\mrm{r}(t)/\left[ I_\mrm{r}(t) + I_\mrm{s}(t) \right]$.
 The rate of its change is obtained from
 Eqs.\,(\ref{eq:1host2fung-gen-1})-(\ref{eq:1host2fung-gen-3})
\begin{equation}\label{eq:resfreq-dyn}
\frac{d p}{d t} = s(t) p (1 - p),
\end{equation}
where 
\begin{equation}\label{eq:sel-coeff}
  s = b \left[ (1 - \veps_\mrm{r}(C, r_\mrm{B})) (1 - \rho_\mrm{r} ) -
    (1 -\veps_\mrm{s}(C, r_\mrm{B})) \right] H(t)
\end{equation}
is the selection coefficient [a similar expression was found in
\cite{gugi99}]. Here $\veps_\mrm{s}(C, r_\mrm{B})$ and
$\veps_\mrm{r}(C, r_\mrm{B})$ are given by \eq{eq:veps-mix-comp-sens}
and \eq{eq:veps-mix-comp-res}.
If $s>0$, then the resistant strain is favored by selection and will
eventually dominate the pathogen population ($p \to 1$ at $t \to
\infty$). Alternatively, if $s<0$, then the sensitive strain is
selected and will dominate the population ($p \to 0$ at $t \to
\infty$). 

The focus of this paper is to investigate the parameter range
over which $s<0$, i.\,e. the sensitive strain is favored by
selection. Mathematically this corresponds to finding the range of
stability of the equilibrium (fixed) point of the system
Eqs.\,(\ref{eq:1host2fung-gen-1})-(\ref{eq:1host2fung-gen-3}),
corresponding to $H>0$, $I_\mrm{s}>0$, $I_\mrm{r}=0$. Our focus is
mainly on the direction of selection. To address this point we
do not need to assume that the host-pathogen equilibrium is
reached. However, we explicitly assume that the host-pathogen
equilibrium is reached during one season in \sec{sec:trben}, where we
evaluate the benefit of fungicide treatment. The implications of this
assumption are discussed at the beginning of \sec{sec:trben}.
Furthermore, we assume that the fungicide dose is constant over
time. (See Appendix A.\,4 for the justification of this assumption.)


A careful examination of the \eq{eq:sel-coeff} reveals that the sign
of the selection coefficient $s$, and therefore the direction of
selection, is determined by the expression in square brackets, which
can be either positive or negative depending on the values of $C$,
$r_\mrm{B}$, $\rho_\mrm{r}$ and the shapes of the functions
$\veps_\mrm{s,r}$. The sign of the selection coefficient is unaffected
by $b$ and $H(t)$ since both of them are non-negative. Consequently
most of the results of this paper do not depend on a particular shape
of $H(t)$ and hence are independent of a particular form of the
growth term (except for those in \sec{sec:trben} concerned with the
benefit of fungicide treatment). This means that the main conclusions
of the paper remain valid for both perennial crops, where the amount of
healthy host tissue steadily increases over many years, and for annual
crops, where the healthy host tissue changes cyclically during
each growing season.

We also neglect the spatial dependencies of the variables $H$,
$I_\mrm{s}$ and $I_\mrm{r}$ and all other parameters. The latent phase
of infection, which can be considerable for some pathogens, is also
neglected.
Since we neglect mutation, migration and spatial heterogeneity, the
resistant and sensitive pathogen strains cannot co-exist in the long
term (see Appendix\,A.1). Only one of them eventually survives: the
one with a higher basic reproductive number.

The basic reproductive number, $R_0$, is often used in epidemiology as
a measure of transmission fitness of infectious pathogens
\cite{anma86}. It is defined as the expected number of secondary
infections resulting from a single infected individual introduced into
a susceptible (healthy) population. At $R_0>1$ the infection can spread over the
population, while at $R_0<1$ the epidemic dies out.

The equilibrium stability analysis of the model
Eqs.\,(\ref{eq:1host2fung-gen-1})-(\ref{eq:1host2fung-gen-3}) (see
Appendix\,A.1) shows that the relationship between the basic
reproductive number of the sensitive strain $R_\mrm{0s} = b \left(1 -
  \veps_\mrm{s}(C, r_\mrm{B}) \right) K/\mu $ and the basic
reproductive number of the resistant strain $R_\mrm{0r} = b \left(1 -
  \veps_\mrm{r}(C, r_\mrm{B}) \right) (1 - \rho_\mrm{r}) K / \mu$
determines the long-term outcome of the epidemic. The sensitive strain
wins the competition and dominates the pathogen population if
$R_\mrm{0s}>1$, such that it can survive in the absence of the
resistant strain, and $R_\mrm{0s} > R_\mrm{0r}$ (this is equivalent to
$s<0$), such that it has a selective advantage over the resistant
strain.  The latter inequality is equivalent to
\begin{align}
\veps_\mrm{s}(C,r_\mrm{B})   < \rho_\mrm{r} + \veps_\mrm{r} (C, r_\mrm{B}) (1 - \rho_\mrm{r}) \label{eq:brn-ineq-gen0}
\end{align}
Similarly, the resistant strain wins the competition and dominates the
population if $R_\mrm{0r}>1$ and $R_\mrm{0r} > R_\mrm{0s}$ (this is equivalent to
$s>0$).

We determined the range of the fungicide doses $C$ and fitness costs
$\rho_\mrm{r}$, according to the inequality (\ref{eq:brn-ineq-gen0})
analytically when (i) the high-risk fungicide has a higher efficacy
than the low-risk fungicide ($k_\mrm{kA}>k_\mrm{kB}$), but
pharmacological interaction is absent ($u=0$); and (ii) the two
fungicides have the same efficacy ($k_\mrm{kA}=k_\mrm{kB}=k_\mrm{k}$),
but may interact pharmacologically ($u \ne 0$).
In case (i) the criterium (\ref{eq:brn-ineq-gen0}) assumes the form
\begin{equation}\label{eq:brn-ineq-gen-kka-kkb}
\frac{ C}{ C + C_{50}}   < \rho_\mrm{r}/k_\mrm{km} + k_\mrm{kB}/k_\mrm{km} \frac{ C}{ C +
    C_{50}/\gamma_\mrm{r}} (1 - \rho_\mrm{r}),
\end{equation}
where $k_\mrm{km} = k_\mrm{kA}(1-r_\mrm{B}) + k_\mrm{kB} r_\mrm{B}$, while in case (ii) the criterium (\ref{eq:brn-ineq-gen0}) reads
\begin{equation}\label{eq:brn-ineq-gen}
\frac{ C}{ C + C_{50}/\gamma_\mrm{s}}   < \rho_\mrm{r}/k_\mrm{k} + \frac{ C}{ C +
    C_{50}/\gamma_\mrm{r}} (1 - \rho_\mrm{r}),
\end{equation}

To keep the presentation concise, below we present the results
corresponding to case (ii), i.\,e. solve the inequality
(\ref{eq:brn-ineq-gen}). However, we verified that all the conclusions remain the same in case (i). 
In a more general case, when $k_\mrm{kA}>k_\mrm{kB}$ and $u \ne 0$ the
parameter ranges satisfying the inequality (\ref{eq:brn-ineq-gen0})
can only be determined numerically.

We assume here that both the cost of resistance and fungicidal
activity decrease the transmission rate $b$. However, we performed the
same analysis when the effect of the resistance cost and the fungicide
enter the model in other ways and obtained qualitatively similar
results (see Appendix\,A.5)

The simplicity of the model allows us to obtain all of the results
analytically. We determined explicit mathematical relationships
between the quantities of interest, which enabled us to study the
effects over the whole range of parameters.

\section{Results}
\label{sec:results}

%
We first investigate the parameter ranges over which resistant or
sensitive strains dominate the pathogen population for the case of
fungicides A and B applied individually and in a mixture
(\sec{sec:selres}). Then, we consider the optimal proportion of
fungicides to include in a mixture in \sec{sec:optrat} and the benefit
of fungicide treatment in \sec{sec:trben}. Finally, we take into
account possible pharmacological interactions between fungicides and
consider the effect of partial resistance (\sec{sec:pharmint},
\ref{sec:partres}).

\subsection{Selection for resistance}
\label{sec:selres}

The ranges of fungicide dose and cost of resistance at which the
sensitive (white) or resistant (grey) pathogen strain is favored by
selection are shown in \fig{fig:phase-diagr}. In all scenarios
competitive exclusion is observed: one of the strains takes over the
whole pathogen population and the other one is eliminated. If a low-risk
fungicide is applied alone, the sensitive strain has a selective
advantage across the whole parameter range in
\fig{fig:phase-diagr}A. When only a high-risk fungicide is applied
(\fig{fig:phase-diagr}B) the resistance dominates if the fitness cost
is lower than the maximum effect of the fungicide
$\rho_\mrm{r}<k_\mrm{k}$ and at a fungicide dose higher than a
threshold value which increases with the fitness cost (solid curve in
\fig{fig:phase-diagr}B). If the fitness cost exceeds $k_\mrm{k}$
(dotted line in \fig{fig:phase-diagr}B), then the sensitive strain
dominates at any fungicide dose.
The \fig{fig:phase-diagr}C shows the outcome when the two fungicides
are mixed at equal concentrations.
Here the fitness cost at which the sensitive strain dominates is
reduced (vertical dotted line is shifted to the left).
%

As expected, without a fitness cost ($\rho_\mrm{r} = 0$) the resistant
strain becomes favored by selection and will eventually dominate the
population whenever the high-risk fungicide is applied, alone or in
combination with the low-risk fungicide (\fig{fig:phase-diagr}B,C).
%

\subsection{Optimal proportion of fungicides in a mixture}
\label{sec:optrat}

It is highly desirable to keep existing fungicides effective for as
long as possible. 
From this point of view, an optimal mixture contains the largest proportion of the
high-risk fungicide, at which (i) the resistant pathogen strain is not
selected and (ii) an adequate level of disease control is achieved.
In order to fulfill both of these objectives, the fitness cost of
resistance needs to be larger than a threshold value $\rho_\mrm{r} >
\rho_\mrm{rb}$ [see Eq.\,(A.16)]. The threshold
$\rho_\mrm{rb}$ is shown by the dotted vertical line in \fig{fig:phase-diagr}C.
%

The threshold $\rho_\mrm{rb}$ depends on the proportion of fungicides
in the mixture.
Adding more of the low-risk fungicide, while keeping the same total
dose $C$, reduces the threshold. This diminishes the range of the
values for fitness cost over which the resistant strain dominates. On
the other hand, adding less of the low-risk fungicide, while again
keeping $C$ the same, increases the threshold, which increases the
parameter range over which the resistant strain is favored.

Therefore, at a given fitness cost $\rho_\mrm{r}$, one can adjust the
fungicide ratio $r_\mrm{B}$ such that
$\rho_\mrm{r}>\rho_\mrm{rb}$. This is shown in
\fig{fig:opt-fung-rat-comb}: the curve shows the critical proportion
of the low-risk fungicide $r_\mrm{Bc}$, above which no selection for
resistance occurs at any total fungicide dose $C$. One can
see from \fig{fig:opt-fung-rat-comb} that if the resistance cost is
absent ($\rho_\mrm{r} = 0$), then the high-risk fungicide should not
be added at all if one wants to prevent selection for resistance.
At larger fitness costs, the value of $r_\mrm{Bc}$ decreases, giving
the possibility to use a larger proportion of the high-risk fungicide
without selecting for resistance.

Finding an optimum proportion of fungicides requires knowledge of both
the fitness cost $\rho_\mrm{r}$ and the maximum effect of the
fungicide $k_\mrm{k}$ [Eq.\,(A.17)]. However, if the cost of
resistance and fungicides affect the infectious period of the pathogen
$\mu^{-1}$ (see Sec.\,A.5) and not the transmission rate $b$ as
we assumed above, then a simpler expression for the critical
proportion of fungicides in the mixture is obtained
[Eq.\,(A.18)], which depends only on the ratio between the
fitness cost and the maximum fungicide effect $\rho_\mrm{r}/k_k$. In
this case, if the fitness cost is at least 5 percent of the maximum
fungicide effect, then we predict that up to about 20 percent of the
high-risk fungicide can be used in a mixture without selecting for
resistance. 
An example of the cost of fungicide resistance manifesting
as a reduction in infectious period was in metalaxyl-resistant
isolates of \emph{Phytophthora infestans} \cite{kaco89}. In this
experiment, the infectious period of the resistant isolates was
reduced, on average, by 25\,\% compared to the susceptible isolates
\cite{kaco89}.

So far we have shown how choosing an optimal proportion of fungicides
in the mixture prevents selection for resistance. Now, we will
consider in more detail how to achieve an adequate level of disease control.

\subsection{Treatment benefit}
\label{sec:trben}

The yield of cereal crops is usually assumed
to be proportional to the healthy green leaf area, which corresponds
in our model to the amount of healthy hosts $H(t)$. Accordingly,
we quantify the benefit of the fungicide treatment, $B(t)$, as the ratio
between the amount of healthy hosts $H(t)$ when both the disease
and treatment are present and its value $H_\mrm{nd}(t)$ in the absence
of disease: $B(t) = H(t)/H_\mrm{nd}(t)$. Hence, $B(t)=1$ corresponds
to a perfect treatment, which eradicates the disease completely and
the treatment benefit of zero corresponds to a situation where all
healthy hosts are infected by disease.
In order to obtain analytical expressions for the treatment benefit
$B(t)$, we consider one growing season and assume that the host-pathogen
equilibrium is reached during the season (see Appendix\,A.3 for a
discussion of these assumptions).

The treatment benefit at equilibrium is shown in
\fig{fig:trben-vs-conc-cost-col} as a function of the fitness cost and
the fungicide dose (see Appendix\,A.3 for equations).
When a low-risk fungicide is applied alone
[\fig{fig:trben-vs-conc-cost-col}A], the sensitive strain is favored
by selection over the whole range of parameters. Therefore, the
treatment benefit increases monotonically with the fungicide dose and
is not affected by the cost of resistance.
In contrast, when a high-risk fungicide is applied alone
[\fig{fig:trben-vs-conc-cost-col}B], a region at low fitness costs appears [to
the left from the solid curve in \fig{fig:trben-vs-conc-cost-col}B],
where the resistant strain is favored. Here, the treatment
benefit does not depend on the fungicide dose, but increases with
the cost of resistance. Hence, if the fitness cost is too low to stop
selection for resistance, then the fungicide treatment will fail.

In the case of a mixture of a high-risk and a low-risk fungicide, the
parameter range over which the resistant strain is favored becomes
smaller [\fig{fig:trben-vs-conc-cost-col}C, to the left from the solid
  curve].
In this range the treatment benefit increases with the cost of resistance,
since larger costs reduce the impact of disease per se. Also, the
treatment benefit increases with the total fungicide dose in this
range, because the low-risk fungicide works against the resistant
strain.


As we have shown above in \sec{sec:optrat}, in the presence of a
substantial fitness cost, one can avoid selection for resistance by
adjusting the proportion of the two fungicides in the mixture. Then,
the total fungicide dose such that the treatment benefit reaches a
high enough value and an adequate disease control is achieved.

In the end of \sec{sec:optrat} we estimated that up to about 20 percent of the
high-risk fungicide can be used in a mixture without selecting for
resistance if the fitness cost is at least 5 percent of the maximum
fungicide effect on the infectious period $\mu^{-1}$.
But how much extra control does one obtain by adding the high-risk
fungicide to the mixture?
We estimate that adding 20 percent of the high-risk fungicide to the mixture
increases the treatment benefit by about 12 percent at $R_\mrm{0s}(C=0)=b
K/\mu=4$ and by about 9 percent at $R_\mrm{0s}(C=0)=2$ (see Sec.\,A.3 and
Fig.\,A.1 for more details). In the case when the high-risk fungicide
has a larger maximum effect, i.\,e. $k_\mrm{kA}>k_\mrm{kB}$, the
benefit of adding it to the mixture will increase. However, the largest
proportion of the high-risk that can be added without selecting for
resistance will decrease.


\subsection{The effect of pharmacological interaction between
  fungicides}
\label{sec:pharmint}
Synergistic interactions between fungicides make their combined effect
greater than expected with additive interactions.  The sensitive
pathogen strain is suppressed more by a synergistic mixture, while the
resistant strain is not affected by the interaction (in case of full
resistance $\alpha=0$). This increases the range of fitness costs over
which resistance has a selective advantage [the dashed line in
\fig{fig:phase-diagr}C shifts to the right]. Consequently, the
critical proportion of the low-risk fungicide in the mixture
$r_\mrm{Bc}$, above which the resistant mutants are eliminated
increases [dotted curve in
\fig{fig:rbc-vs-rhoa-interact-partres}A]. In contrast, an antagonistic
mixture suppresses the sensitive strain less effectively than either
fungicide used alone. In this case the range of fitness costs over
which resistance dominates becomes smaller and the ratio $r_\mrm{Bc}$
decreases [dashed curve in
\fig{fig:rbc-vs-rhoa-interact-partres}A]. Hence, reduced resistance
evolution is achieved, however, at the expense of reduced disease
control.
This result is in agreement with studies on drug interactions in the
context of antibiotic resistance, where antagonistic drug combinations
were found to select against resistant bacterial strains \cite{chcr+07}.

\subsection{The effect of partial fungicide resistance}
\label{sec:partres}
Consider the situation when the resistant pathogen strain is not fully
protected from the high-risk fungicide, but exhibits a partial
resistance ($0 < \alpha < 1$). In this case, the fungicide mixture is
more effective in suppressing the resistant strain than in the case of
full resistance ($\alpha=0$) considered above.  Therefore, one needs
less of the low-risk fungicide in the mixture to reach the conditions
where resistance is eliminated by selection: the critical proportion
of the low-risk fungicide in the mixture decreases with the degree of
sensitivity $\alpha$ in
\fig{fig:rbc-vs-rhoa-interact-partres}B. Also, in
\fig{fig:rbc-vs-rhoa-interact-partres}A the dependency of the
critical ratio of the fungicide B in the mixture for partial
resistance (light grey curve) lies below the one at perfect resistance
and reaches zero at a much smaller value.
Thus, knowledge of the degree of resistance is crucial for
determining an appropriate proportion of fungicides in the mixture.

\section{Discussion}
\label{sec:conclusions}

The three main outcomes of our study are: (i) if fungicide resistance
comes without a fitness cost, application of fungicides prone to
resistance (high-risk fungicides) in a mixture with fungicides still
free from resistance (low-risk fungicides) will select for resistance;
(ii) if sufficiently high costs are found, then an optimal proportion
of the high-risk fungicide in a mixture with the low-risk fungicide
exists that does not select for resistance; (iii) this mixture can
potentially be used for preventing de novo emergence of fungicide
resistance, in which case the relevant fitness cost is the
``inherent'' cost of fungicide resistance before the compensatory
evolution occurs (see below).

In the absence of fitness costs application of a mixture of high-risk
and low-risk fungicides will select for resistance.
Consequently, the resistant strain will eventually dominate the pathogen
population and the sensitive strain will be eliminated. Because of
this, the high-risk fungicide will not affect the amount of disease
and only the low-risk fungicide component of the mixture will be
acting against disease.
Hence, the high-risk fungicide becomes nonfunctional in the mixture
and using the low-risk fungicide alone would have the same effect at a
lower financial and environmental cost.
In contrast, if sufficiently high costs are found, then high-risk
fungicides can be used effectively for an extended period of
time. According to our model, an optimal proportion of the high-risk
fungicide in a mixture with the low-risk fungicide can be determined
that contains as much as possible of the high-risk fungicide, but
still does not select for resistance while providing adequate disease
control (see Box\,\ref{nbox:getrb-alg}). If a mixture with the optimum
proportion is applied, then the rise of the resistant strain is
prevented for an unlimited time. Thus, the scheme in
Box\,\ref{nbox:getrb-alg} provides a framework for using our knowledge
about the evolutionary dynamics of plant pathogens and their
interaction with fungicides in devising practical strategies for
management of fungicide resistance.

In order to apply the scheme in Box\,\ref{nbox:getrb-alg}, one needs
to know dose-response parameters of the fungicides $k_\mrm{k}$ and
$C_{50}$, the degree of fungicide sensitivity $\alpha$ (or the
resistance factor), the degree of pharmacological interaction $u$ and
the fitness cost of resistane mutations.
%
%
%
Fungicide dose-response curves are routinely determined empirically
(for example, \cite{locl05,pahi+98}) and can be used to estimate the
model parameters $k_\mrm{k}$ and $C_{50}$ \cite{hopa+11a}. The
fungicide sensitivity is known to be lost completely in some cases
(for example, most cases of QoI resistance), i.\,e. $\alpha=0$, while
in other cases with partial resistance the degree of sensitivity (or
the resistance factor) was
measured (for example, \cite{leal+07}). Pharmacological interaction between several different
fungicides was also characterized empirically (see \cite{gi96} and the
references therein). Also, the fitness costs of resistance were
characterized empirically in many cases (see below).
%
In the past these measurements were performed independently, but our
study provides motivation to bring them together, since all these
parameters need to be characterized for the same
plant-pathogen-fungicide combination.

These measurements will allow one to predict the optimal proportion of the
two fungicides in the mixture theoretically. This prediction needs to
be tested using field experimentation, in which the amount of disease
and the frequency of resistance would be measured as functions of time
at different proportions of the high- and low-risk fungicides in the
mixture.
 From these measurements the optimal proportion of the fungicides can
 be obtained empirically. It is this empirically determined optimal
 proportion of fungicides that can be used for practical guidance on
 management of fungicide resistance. Moreover, from the comparison of
 the optimal proportion obtained theoretically and empirically, one
 can evaluate the performance of the model and identify the aspects of
 the model that need improvement.



 So far we considered the scenario where both the sensitive and the
 resistant pathogen strains increase from low numbers,
 i.\,e. resistant mutants pre-exist in the pathogen population. In
 this scenario the strain with higher fitness (or basic reproductive
 number) eventually outcompetes the other strain. This competition may
 occur over a time scale of several growing seasons so that there is
 enough time for compensatory mutations that diminish fitness costs of
 resistance to emerge. This needs to be taken into account when
 determining the optimal proportion of fungicides in the
 mixture. However, an alternative scenario is possible when resistant
 mutants emerge de novo through mutation or migration and, in order to
 survive, they need to invade the host population already infected by
 the sensitive strain. The threshold of invasion in this case depends on the
 ``inherent'' fitness cost of resistance mutations, i.\,e. their cost
 before the compensatory evolution occurs. In this case, one should
 measure the ``inherent'' cost of resistance mutations when performing
 step 3 in Box\,\ref{nbox:getrb-alg}.

As discussed above, it is crucial to know fitness costs of
resistance mutations in order to determine whether the fungicide
mixture will select for resistance. We extensively searched the
literature on fitness costs in different fungal pathogens of plants.
A few studies inferred substantial fitness costs from field monitoring
(see for example, \cite{suya+10} and references in \cite{pemi95}). But
these findings could result from other factors, including immigration
of sensitive isolates, selection for other traits linked to resistance
mutations or genetic drift \cite{pemi95}. Though relatively few
carefully controlled experiments have been conducted, the majority
indicate that fitness costs associated with fungicide resistance are
either low (for example, \cite{bifi+12,kixi11}) or absent (for
example, \cite{codu+10,pemi94}). But in some cases fitness costs were
found to be substantial (for example,
\cite{iaba+08,kath+01,hoec95,we88}) both in laboratory measurements
and in field experiments.
Although measurements of fitness costs of resistant mutants performed
under laboratory conditions can be informative (as, for example in
\cite{bifi+12}), they do not necessarily reflect the costs connected
with resistant mutants selected in the field. This is because field
mutants are likely to possess compensatory mutations improving
pathogen fitness \cite{pemi95}. Moreover, a laboratory setting rarely
reflects the balance of environmental and host conditions found
throughout the pathogen life-cycle, since the field environment is
much more complex.


However, the most relevant measure of pathogen fitness in the context
of our study is the growth rate of the pathogen population at the very
beginning of an epidemic (often denoted as $r$). It is directly
related to the basic reproductive number $R_0$.  To the best of our
knowledge, the fitness costs of fungicide resistant strains were not
measured with respect to $r$. In the studies cited above different
components of fitness were measured that may or may not be related to
$r$.
Therefore, we identified a major gap in our knowledge of fitness costs. We
hope this study will stimulate further experimental investigations to
better characterize fitness costs and expect that substantial costs
will be found in some cases.

Interactions of plants with fungal pathogens, fungicide action and,
possibly, pharmacological interaction can depend on environmental
conditions. This means that the outcomes of measurements necessary for
applying the scheme in Box\,\ref{nbox:getrb-alg}, may vary between
seasons and geographical locations. Moreover, the outcomes may also be
different in different host cultivars. Therefore, the optimal
proportion of fungicides in the mixture may vary between seasons,
geographical locations and host cultivars. Thus, to provide general
practical guidance on management of fungicide resistance, one needs to
measure the optimal proportion of fungicides over many seasons, in
different geographical locations and host cultivars. This difficulty
is not a unique property of our study, but rather it is a general
problem in the field of mathematical modeling of fungicide resistance
and plant diseases. For example, it is also relevant for choosing
appropriate fungicide dose rate \cite{locl05}.


While it was previously discussed \cite{sh06} that alternation of
high-risk and low-risk fungicides might be a useful tactic for disease
control in the presence of a fitness cost, we have shown that a
mixture of these fungicides in an appropriate proportion can provide
adequate disease control without selecting for resistance. Mixtures
offer an advantage compared to alternation because there is no need to
delay the application of the high-risk fungicide and the resistant
strain does not rise to high frequencies, which lowers the risk of its
further spread (see Appendix\,A.6).

\begin{nicebox}
According to our model, one can avoid selection for resistance while providing adequate
disease control by choosing the fungicide ratio $r_\mrm{B}$ and the
total dose $C$ in the following way:
\begin{enumerate}
\item measure the pharmacological properties of both fungicides under field conditions to
  determine $k_\mrm{k}$, and $C_{50}$;
\item determine the degree of fungicide sensitivity $\alpha$ under field conditions;
\item determine the degree of pharmacological interaction $u$ between
  fungicides A and B under field conditions;
\item measure the fitness cost of resistance $\rho_\mrm{r}$ under field conditions;
\item choose the proportion of the fungicide B above the threshold:
$r_\mrm{B}>r_\mrm{Bc}$, such that the resistance is not favored by selection at any
total fungicide dose $C$;
\item choose the total fungicide dose, which should be large enough to
  achieve an adequate  level of disease control (see \fig{fig:trben-vs-conc-cost-col}C).
\end{enumerate}
\caption{How to determine an optimal mixture of fungicides theoretically.}\label{nbox:getrb-alg}
\end{nicebox}


The problem of combining chemical biocides in order to delay or
prevent the development of resistance also appears in other contexts,
including resistance of agricultural weeds to herbicides \cite{bere09} and
insect pests to insecticides \cite{ro06}.
The fitness cost of resistance is also recognized as a crucial
parameter for managing antibiotic resistance \cite{anhu10}. 


%
Development of mathematical models of fungicide resistance dynamics
has been influenced by theoretical insights from animal and human
epidemiology \cite{bogi08,hagu+04}. Similarly, we expect that lessons
learned from modeling fungicide combinations may well apply to the
problem of biocide resistance in the other contexts. In particular,
one can investigate the idea of adjusting the proportion of the
components in a mixture of drugs in order to prevent selection for
resistance in a more general context of biocide resistance.

\section*{Acknowledgements}

AM and SB gratefully acknowledge support by the European Research
Council advanced grant PBDR 268540. The authors are grateful to
Michael Milgroom and Michael Shaw for helpful comments concerning
fitness costs of fungicide resistance and to two anonymous reviewers
for improving the manuscript.

\newpage

\clearpage
\begin{figure}[!ht]
  \centerline{\includegraphics[width=\textwidth]{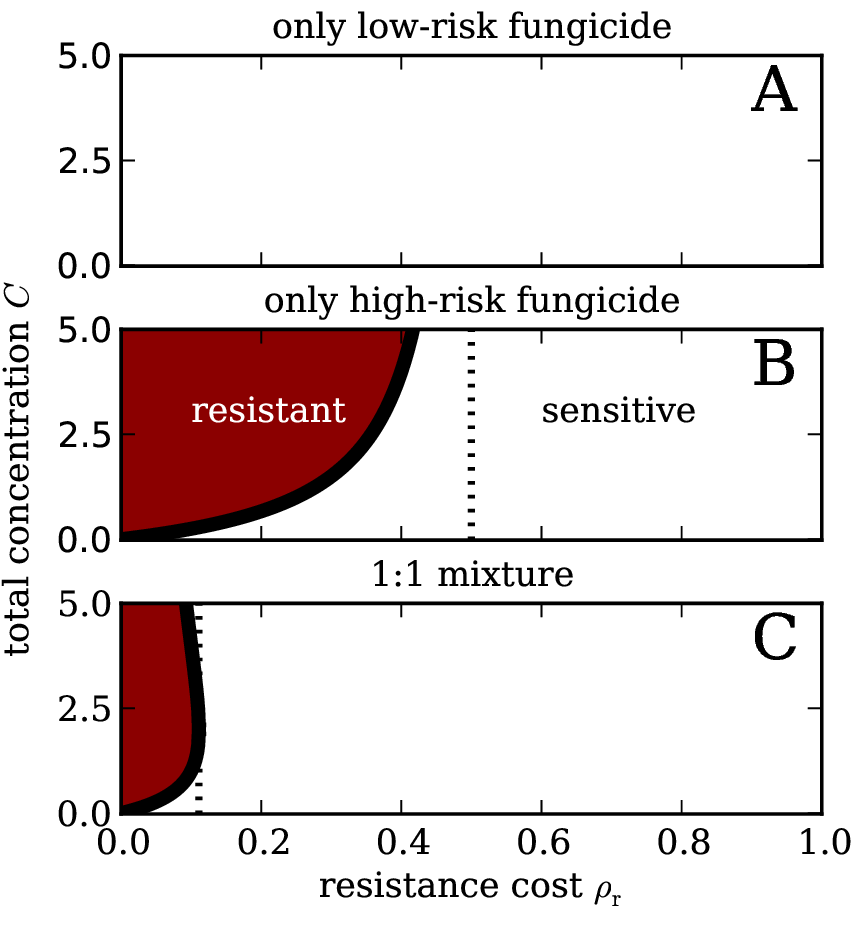}} \caption{
    \doublespacing 
Outcomes of the competition between the sensitive and resistant
    pathogen strains depending on the fitness cost of resistance
    $\rho_\mrm{r}$ and the fungicide dose $C$ when treated
    with a single fungicide B at [$C_\mrm{B} = C$, panel A], a
    single fungicide A [$C_\mrm{A}=C$, panel B] and the combination
    of fungicides A and B [$C_\mrm{A} = C_\mrm{B} = C/2$, panel
    C]. The range of the total fungicide dose $C$ and the
    fitness cost of resistance $\rho_\mrm{r}$, in which the resistant
    strain is favored is shown in grey. The range where selection
    favors the sensitive strain is shown in white. The dashed and the
    solid curves in panel B are plotted according to Eq.\,(A.22) and
    Eq.\,(A.23) in Appendix\,A.2, respectively. The dashed and the
    solid curves in panel C are plotted according to Eq.\,(A.13) and
    Eq.\,(A.14) in Appendix\,A.1, respectively, at $\gamma_\mrm{s}=1$,
    $\gamma_\mrm{r}=1/2$. Fungicides are assumed to have zero
    interaction ($u = 0$) and the resistant strain is assumed to be
    fully protected from fungicide A ($\alpha=0$), the fungicide
    dose-response parameters are $k_\mrm{k}=0.5$, $C_{50}=1$.}
\label{fig:phase-diagr}
\end{figure}%
\clearpage
\begin{figure}[!ht]
  \centerline{\includegraphics[width=0.6\textwidth]{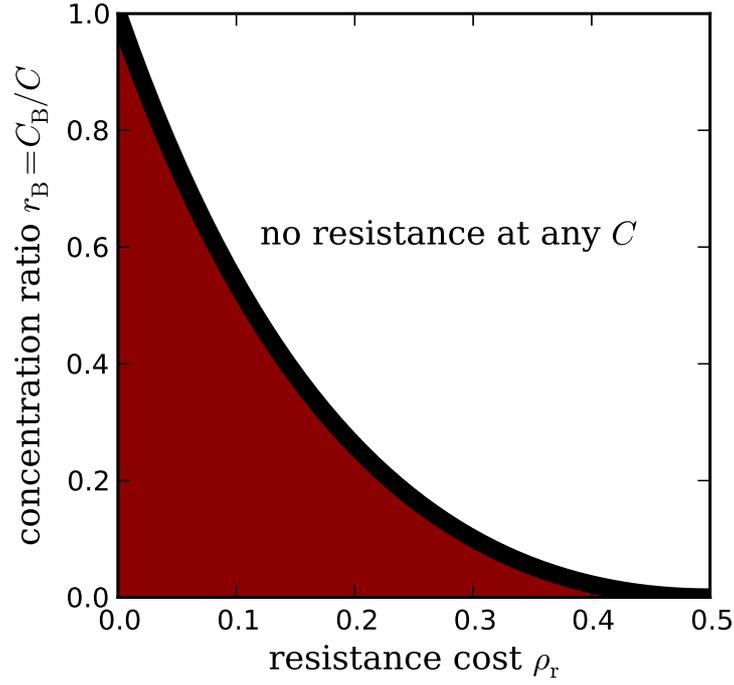}} \caption{
    \doublespacing The critical proportion $r_\mrm{Bc}$ of fungicide B
    (low-risk fungicide) in the mixture, above which there is no
    selection for the resistant strain at any total fungicide
    dose $C$, plotted (black curve) according to Eq.\,(A.17)
    as a function of the resistance cost $\rho_\mrm{r}$, assuming no
    pharmacological interaction ($u=0$), full resistance ($\alpha=0$)
    and the maximum fungicide effect $k_\mrm{k}=0.5$.}
\label{fig:opt-fung-rat-comb}
\end{figure}%
\clearpage
\begin{figure}[!ht]
  \centerline{\includegraphics[width=0.6\textwidth]{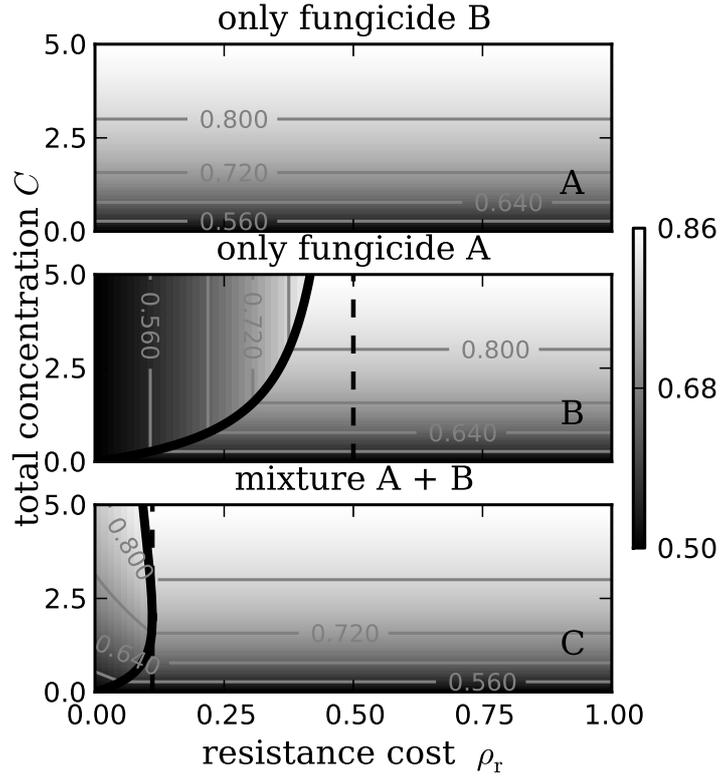}} \caption{
\doublespacing 
    Treatment benefit as a function of fungicide dose $C$ and
    fitness cost of resistance $\rho_\mrm{r}$, plotted according to
    Eq.\,(A.26) in panel A, according to Eq.\,(A.27) in panel B
    and according to Eq.\,(A.28) in panel C. Treatment with
    fungicide B is shown in panel A. Treatment with fungicide A is
    shown in panel B. Treatment with a mixture of A and B at equal
    concentrations ($r_\mrm{B}=1/2$) is shown in panel C. Solid and
    dashed curves in panels B and C are the same as in
    \fig{fig:phase-diagr}.  Fungicides are assumed to have zero
    interaction ($u = 0$) and the resistant strain is assumed to be
    fully protected from fungicide A ($\alpha=0$). The fungicide
    dose-response parameters are $k_\mrm{k}=0.5$, $C_{50}=1$, the basic
    reproductive number of the sensitive strain without fungicide
    treatment $R_\mrm{0s}(C=0) =b K / \mu=2$.}
\label{fig:trben-vs-conc-cost-col}
\end{figure}%
\clearpage
\begin{figure}[!ht]
  \centerline{\includegraphics[width=0.8\textwidth]{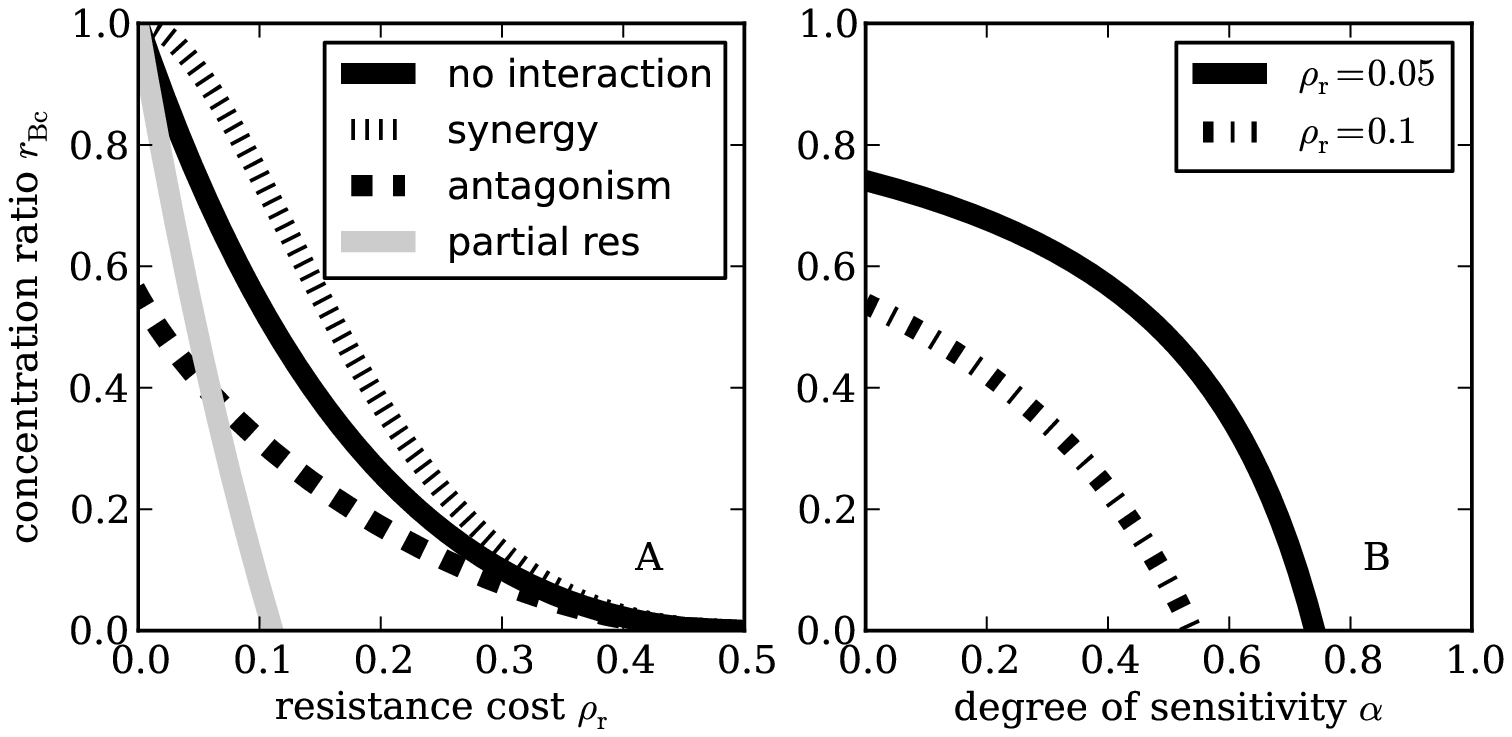}} \caption{
\doublespacing 
    The effect of pharmacological interaction and partial resistance
    on $r_\mrm{Bc}$, the critical ratio of the fungicide
    B. $r_\mrm{Bc}$ is plotted as a function of the fitness cost of
    resistance $\rho_\mrm{r}$ (left panel), according to
    Eq.\,(A.13) for the case of no interaction between the
    fungicides $u=0$ (solid, the same as the curve in
    \fig{fig:opt-fung-rat-comb}), synergy $u=0.9$ (dotted), and
    antagonism $u=-0.9$ (dashed) for the case of perfect resistance
    $\alpha=1$. The case of partial resistance at no interaction
    ($\alpha = 0.5$, $u=0$) is shown as a light grey curve.
    $r_\mrm{Bc}$ is shown as a function of the degree of fungicide
    sensitivity $\alpha$ at $\rho_\mrm{r}=0.05$ (solid) and
    $\rho_\mrm{r}=0.1$ (dash-dotted) also according to
    Eq.\,(A.13) in the right panel.}
\label{fig:rbc-vs-rhoa-interact-partres}
\end{figure}%
\newpage

\bibliography{/Users/alexey/Dropbox/mybib/evolut}
\bibliographystyle{phytopath}


\doublespacing

\appendix
\numberwithin{equation}{section}
\renewcommand\thefigure{\thesection.\arabic{figure}}    
\setcounter{figure}{0}  


\section{Supplemental materials}

\subsection{Model equations}
\label{sec:ap-modeleq}

In order to explore the effect of the assumptions we made in
\sec{sec:model-assump}, we consider a more general system of
equations, which describes the change in time of the same quantities
as in Eqs.\,(\ref{eq:1host2fung-gen-1})-(\ref{eq:1host2fung-gen-3}):
the amount of healthy host tissue $H$, the amount of host tissue
infected with the sensitive pathogen strain $I_\mrm{s}$ and the amount
of host tissue infected with the resistant pathogen strain $I_\mrm{r}$

%
\begin{align}
\frac{d H}{d t} &= r_H (K - H - I_\mrm{s} - I_\mrm{r} ) -  b \left[
  \left( 1 - \veps_\mrm{s} \right)
(1-\rho_\mrm{s}) I_\mrm{s} + \left( 1 - \veps_\mrm{r} \right)  (1-\rho_\mrm{r}) I_\mrm{r} \right] H, \label{eq:1host2fung-gen-app-1} \\
\frac{d I_\mrm{s}}{d t} & = b \left( 1 - \veps_\mrm{s} \right) (1-\rho_\mrm{s}) H I_\mrm{s} - \mu  I_\mrm{s}, \label{eq:1host2fung-gen-app-2}\\
\frac{d I_\mrm{r}}{d t} & = b \left( 1 - \veps_\mrm{r} \right) (1-\rho_\mrm{r}) H I_\mrm{r} - \mu  I_\mrm{r}, \label{eq:1host2fung-gen-app-3}
\end{align}
where, the function $\veps_\mrm{s} = \veps_\mrm{s}(C_\mrm{A}, C_\mrm{B})$ describes the effect of the
application of the mixture fungicides A and B with doses
$C_\mrm{A}$ and $C_\mrm{B}$ on the transmission rate of the sensitive pathogen strain
and the function $\veps_\mrm{r} = \veps_\mrm{r}(C_\mrm{A}, C_\mrm{B})$ describes the effect of this
mixture on the transmission rate of the resistant strain:
\begin{align}
  \veps_\mrm{s}(C_\mrm{A}, C_\mrm{B}) &= k_\mrm{k} \frac{\alpha_\mrm{s,A} C_\mrm{A} + \alpha_\mrm{s,B} C_\mrm{B}}{ \alpha_\mrm{s,A} C_\mrm{A} + \alpha_\mrm{s,B} C_\mrm{B} + C_{50}/ \left[1
    + u \sqrt{ \alpha_\mrm{s,A} \alpha_\mrm{s,B}  C_\mrm{A} C_\mrm{B}} /
    (\alpha_\mrm{s,A} C_\mrm{A} + \alpha_\mrm{s,B} C_\mrm{B})\right]}, \label{eq:veps-s-mix-gen-app}\\
  \veps_\mrm{r}(C_\mrm{A}, C_\mrm{B}) &= k_\mrm{k} \frac{\alpha_\mrm{r,A} C_\mrm{A} + \alpha_\mrm{r,B} C_\mrm{B}}{ \alpha_\mrm{r,A} C_\mrm{A} + \alpha_\mrm{r,B} C_\mrm{B} + C_{50}/ \left[1
    + u \sqrt{ \alpha_\mrm{r,A} \alpha_\mrm{r,B}  C_\mrm{A} C_\mrm{B}} / (\alpha_\mrm{r,A} C_\mrm{A} + \alpha_\mrm{r,B} C_\mrm{B})\right]}. \label{eq:veps-r-mix-gen-app}
\end{align}
The parameters $\alpha_\mrm{s,A}$, $\alpha_\mrm{s,B}$,
$\alpha_\mrm{r,A}$ and $\alpha_\mrm{r,B}$ characterize the degree of
sensitivity of each of the two pathogen strains (index "s" for
the sensitive strain, index ``r'' for the resistant strain) to each of the two
fungicides A and B. Their values are between zero and one. In this
general case both pathogen strains are partially resistant to both
fungicides. The maximum effect of the fungicide is characterized by
the parameter $k_k$ and assumed to be the same for both fungicides.

The parameter $C_{50}$ in Eqs.\,(\ref{eq:veps-s-mix-gen-app}),
(\ref{eq:veps-r-mix-gen-app}) is modified due to pharmacological
interaction between fungicides characterized by the degree of
interaction $u$. At $u=0$ fungicides do not interact, $u>0$ represents
synergy and $u<0$ corresponds to antagonism. [We restrict our
consideration to $u>-1$, since otherwise the term in the square
brackets of Eqs.\,(\ref{eq:veps-s-mix-gen-app}),
(\ref{eq:veps-r-mix-gen-app}) may become negative, which makes no
sense.]  This way to define pharmacological interaction between
compounds is called ``Loewe additivity'' or ``concentration addition''
in the literature \cite{grbr95,be89}. In this approach an interaction
of a compound with itself is set by definition to be additive (zero
interaction). For example, when the fungicide A is mixed with itself,
the resulting sham mixture is neither synergistic, nor antagonistic
but has zero interaction. An equivalent graphic procedure is known as
the Wadley method in the fungicide literature [second method described
in \cite{lebe+86}].

An alternative way to define pharmacological interaction assumes that
the two compounds have independent modes of action and is called
``Bliss independence'' \cite{bl39} or Abbott's formula
\cite{ab25}. However, in this definition a compound can have a
pharmacological interaction with itself, i.\,e. be synergistic or
antagonistic.
The study \cite{sh89a} discusses the definition of ``independent action''
of the two fungicides, according to which the two fungicides are
independent when one fungicide does not affect the evolution of
resistance in the other. According to \cite{sh89a,sh93}, this is only
possible when each of the fungicides affects different stages of the
pathogen life cycle.

There are several ways to introduce a deviation from the zero
interaction regime, in which usually an interaction term is added to
the isobologram equation \cite{grbr95}. We have chosen a specific form
of the interaction term, which is proportional to the square root of
the product of the concentrations of the two compounds [Eq.\,(28) in
\cite{grbr95}]. This form allows for a simple analytical expression of
the effect of the combination in Eqs.\,(\ref{eq:veps-s-mix-gen-app}), (\ref{eq:veps-r-mix-gen-app}).

We assume that the cost of resistance decreases the transmission
rate $b$ by a fixed amount $\rho_\mrm{s}$ for the sensitive strain and
by $\rho_\mrm{r}$ for the resistant strain in
Eqs.\,(\ref{eq:1host2fung-gen-app-1})-(\ref{eq:1host2fung-gen-app-3}).
We restrict our consideration here to the case when the ``sensitive''
pathogen strain is fully sensitive to both fungicides
($\alpha_\mrm{s,A} = \alpha_\mrm{s,B} = 1$) and the ``resistant''
strain can have varying degrees of resistance to the fungicide A
($\alpha_\mrm{r,A} \equiv \alpha$, $0 \le \alpha \le 1$), but is fully
sensitive to the fungicide B ($\alpha_\mrm{r,B} = 1$). Therefore, the
cost of resistance for the sensitive strain is zero $\rho_\mrm{s}
=0$. Then, the fungicide dose-response functions become simpler.

In order to determine the range of fitness costs $\rho_\mrm{r}$ and
fungicide doses $C$, over which the sensitive or resistant strain is
favored by selection, we perform the linear stability analysis of the
fixed points of the system
Eqs.\,(\ref{eq:1host2fung-gen-app-1})-(\ref{eq:1host2fung-gen-app-3}). Fixed
points are the values of $H$, $I_\mrm{s}$ and $I_\mrm{r}$ at which the
expressions on the right-hand side of
Eqs.\,(\ref{eq:1host2fung-gen-app-1})-(\ref{eq:1host2fung-gen-app-3})
equal zero.
The system
Eqs.\,(\ref{eq:1host2fung-gen-app-1})-(\ref{eq:1host2fung-gen-app-3})
has three fixed points: (i) $H^* = K$, $I_\mrm{s}=I_\mrm{r}=0$; (ii)
$H^*= \mu/b_\mrm{s}$, $I_\mrm{s}=r_H (b_\mrm{s} K - \mu)/\left[ b_\mrm{s}
  (\mu+r_H) \right]$, $I_\mrm{r} = 0$;
(iii) $H^*= \mu/b_\mrm{r}$, $I_\mrm{s}= 0 $, $I_\mrm{r} = r_H (b_\mrm{r} K - \mu)/\left[ b_\mrm{r}
  (\mu+r_H) \right]$. Here $b_\mrm{s} = b \left[ 1 -
  \veps_\mrm{s}(C_\mrm{A}, C_\mrm{B})\right]$, $b_\mrm{r} = \left[ 1  - \veps_\mrm{r}(C_\mrm{A},
    C_\mrm{B}) \right] (1 - \rho_\mrm{r})$.
To determine whether a fixed point is stable, we first linearize the system
Eqs.\,(\ref{eq:1host2fung-gen-app-1})-(\ref{eq:1host2fung-gen-app-3})
in its vicinity, then determine the Jacobian and its eigenvalues. A
fixed point is stable if all the eigenvalues have negative real parts.

The results of this analysis can be conveniently expressed using the
basic reproductive number of the sensitive strain
\begin{equation}\label{eq:ap-r0s}
R_\mrm{0s} = \frac{b \left[ 1 - \veps_\mrm{s}(C_\mrm{A}, C_\mrm{B})\right] K}{\mu}
\end{equation}
and the basic reproductive number of the resistant strain
\begin{equation}\label{eq:ap-r0r}
R_\mrm{0r} = \frac{b \left[ 1  - \veps_\mrm{r}(C_\mrm{A},
    C_\mrm{B}) \right] (1 - \rho_\mrm{r}) K}{ \mu}.
\end{equation}
The sensitive strain is favored by selection [meaning that the fixed
point (ii) is stable and both fixed points (i) and (iii) are unstable]
when both inequalities $R_\mrm{0s}>1$, $R_\mrm{0s} > R_\mrm{0r}$ are
fulfilled.

%
%
We consider then the inequality $R_\mrm{0s} > R_\mrm{0r}$, which
is equivalent to
\begin{equation}\label{eq:brn-ineq-gen-app}
\frac{ C}{ C + C_{50}/\gamma_\mrm{s}}   < \rho_\mrm{r}/k_\mrm{k} + \frac{ C}{ C +
    C_{50}/\gamma_\mrm{r}} (1 - \rho_\mrm{r}),
\end{equation}
where
\begin{equation}\label{eq:app-gammas}
\gamma_\mrm{s} = 1 + u \sqrt{r_\mrm{B}(1-r_\mrm{B})},
\end{equation}
\begin{equation}\label{eq:app-gammar}
  \gamma_\mrm{r} = \alpha (1 - r_\mrm{B}) + r_\mrm{B} + u
  \sqrt{\alpha r_\mrm{B}(1-r_\mrm{B})}
\end{equation}
and $r_\mrm{B} = C_\mrm{B}/C$ is the proportion of the funcigide B in
the mixture, $C = C_\mrm{A} + C_\mrm{B}$. 

The inequality (\ref{eq:brn-ineq-gen-app}) holds at
\begin{align}\label{eq:range-sens-gen1}
\rho_\mrm{r} < \rho_\mrm{rb}, \:\: \mrm{for} \left( C<C_\mrm{b1} \: \mrm{or} \: C>C_\mrm{b2}
\right)
\end{align}
or at 
\begin{align}\label{eq:range-sens-gen2}
\rho_\mrm{r} > \rho_\mrm{rb}, \:\: \mrm{for\:any\:value\:of}\: C.
\end{align}
 Here,
\begin{equation}\label{eq:rhorb-gen-app}
\rho_\mrm{rb} = \frac{k_\mrm{k} (\gamma_\mrm{s} - \gamma_\mrm{r}) \left( \gamma_\mrm{s} +
    \gamma_\mrm{r} (1 -k_\mrm{k} )  - 2 \sqrt{\gamma_\mrm{r} \gamma_\mrm{s} (1 - k_\mrm{k})}
  \right) } { \left( \gamma_\mrm{s} - \gamma_\mrm{r} (1 - k_\mrm{k}) \right)^2},
\end{equation}
\begin{equation}\label{eq:c12-gen-app}
C_{b1,2} = \frac{C_{50}}{2 \gamma_\mrm{s} \gamma_\mrm{r} \rho_\mrm{r} (1 - k_\mrm{k})} \left[
  \gamma_\mrm{s} ( k_\mrm{k} - \rho_\mrm{r}) - \gamma_\mrm{r}
  (k_\mrm{k} + \rho_\mrm{r} (1 -  k_\mrm{k})) \mp \sqrt{D} \right],
\end{equation}
where
\begin{equation}\label{eq:d-gen-app}
D = \gamma_\mrm{s}^2 \left( k_\mrm{k} - \rho_\mrm{r} \right)^2 + \gamma_\mrm{r}^2
(k_\mrm{k} -  \rho_\mrm{r} - k_\mrm{k} \rho_\mrm{r} )^2 - 2 \gamma_\mrm{s} \gamma_\mrm{r} \left( k_\mrm{k}^2 (1 - \rho_\mrm{r}) +
  \rho_\mrm{r}^2(1 - k_\mrm{k})  \right).
\end{equation}
According to the inequality (\ref{eq:range-sens-gen2}), if the fitness
cost of resistance is larger than a threshold value given by
\eq{eq:rhorb-gen-app}, the sensitive strain has a selective advantage
and the resistant strain is eliminated from the population at any
fungicide dose $C \geq 0$.


For the case of no interaction between fungicides ($u=0$) and perfect
resistance ($\alpha=0$) we obtain from \eq{eq:app-gammas} and
\eq{eq:app-gammar} $\gamma_\mrm{s} = 1$, $\gamma_\mrm{r} =
r_\mrm{B}$. Then, the \eq{eq:rhorb-gen-app} is simplified:
\begin{equation}\label{eq:rhorb-noint-app}
\rho_\mrm{rb} = k_\mrm{k} \frac{(1 - r_\mrm{B}) \left[ 1 + r_\mrm{B} (1-k_\mrm{k})  - 2
  \sqrt{r_\mrm{B} ( 1 -  k_\mrm{k})} \right]} { (1 -  r_\mrm{B}  (1 -
  k_\mrm{k}) )^2 }.
\end{equation}
We then solve the inequality $\rho_\mrm{r} > \rho_\mrm{rb}$ with respect to
$r_\mrm{B}$ and find that it is fulfilled at $r_\mrm{B}>r_\mrm{Bc}$, where
\begin{equation}\label{eq:rbc-fullres}
r_\mrm{Bc} = \frac{ k_\mrm{k}^2 (1 - \rho_\mrm{r}) + \rho_\mrm{r}^2 ( 1 -
  k_\mrm{k}) -  2 k_\mrm{k} \rho_\mrm{r} \sqrt{ (1 - k_\mrm{k}) (1 - \rho_\mrm{r}) })} { (k_\mrm{k}
  + \rho_\mrm{r} - k_\mrm{k} \rho_\mrm{r})^2}
\end{equation}
It represents the critical proportion of the fungicide B in the
mixture above which the resistant strain is not favored by selection
(\fig{fig:opt-fung-rat-comb}). If the cost of resistance affects the
death rate of the pathogen $\mu$ (see \sec{sec:gener}) and not the
transmission rate $b$ as considered above, then a simpler expression
for the critical proportion of the fungicide B is obtained
\begin{equation}\label{eq:rbc-fullres1}
r_\mrm{Bc} = \left( \frac{1 - \rho_\mrm{r}/k_k}{1 + \rho_\mrm{r}/k_k} \right)^2.
\end{equation}
Here, $r_\mrm{Bc}$ depends only on the ratio $\rho_\mrm{r}/k_k$ of the
cost of resistance to the maximum fungicide effect $k_k$, which allows to
make a more general prediction about the value of $r_\mrm{Bc}$.

\subsection{Selection for resistance at no interaction between fungicides}
\label{sec:app-single-fung}

When only the high risk fungicide (fungicide A) is applied with the
dose $C_\mrm{A}$, we set $r_\mrm{B}=0$ in \eq{eq:app-gammas}
and \eq{eq:app-gammar} to obtain $\gamma_\mrm{s} = 1$, $\gamma_\mrm{r}
= \alpha$. Then, the following expressions are obtained for the
threshold value of the resistance cost from \eq{eq:rhorb-gen-app}
\begin{equation}\label{eq:rhorb-fungica}
\rho_\mrm{rb} =  k_\mrm{k} \frac{(1 - \alpha) \left[ 1 + \alpha (1-k_\mrm{k})  - 2
  \sqrt{\alpha ( 1 -  k_\mrm{k})} \right]} { (1 -  \alpha  (1 - k_\mrm{k}) )^2 },
\end{equation}
and the fungicide dose from \eq{eq:c12-gen-app}
\begin{equation}\label{eq:c12-fungica}
C_{b1,2} = \frac{C_{50}}{2 \alpha \rho_\mrm{r} (1 - k_\mrm{k})} \left[
  k_\mrm{k} - \rho_\mrm{r} - \alpha
  (k_\mrm{k} + \rho_\mrm{r} (1 -  k_\mrm{k})) \mp \sqrt{D} \right],
\end{equation}
where
\begin{equation}\label{eq:d-fungica}
  D = \left( k_\mrm{k} - \rho_\mrm{r} \right)^2 + \alpha^2
(k_\mrm{k} -  \rho_\mrm{r} - k_\mrm{k} \rho_\mrm{r} )^2 - 2 \alpha \left( k_\mrm{k}^2 (1 - \rho_\mrm{r}) +
  \rho_\mrm{r}^2(1 - k_\mrm{k})  \right).
\end{equation}
In the simpler case of full resistance we take the limit $\alpha \to
0$. Then, by taking this limit in \eq{eq:rhorb-fungica},
\eq{eq:c12-fungica} and \eq{eq:d-fungica} we obtain for the threshold
values of the fitness cost and the fungicide dose
\begin{equation}\label{eq:rhorb-fungica-perfres}
\rho_\mrm{rb} = k_\mrm{k},
\end{equation}
\begin{equation}\label{eq:cb-fungica-perfres}
C_\mrm{b} = C_{50} \frac{\rho_\mrm{r}}{k_\mrm{k}- \rho_\mrm{r}}.
\end{equation}
In this case the sensitive strain dominates at $C<C_\mrm{b}$ if $\rho_\mrm{r} <
\rho_\mrm{rb}$ or at any positive values of $C$ if $\rho_\mrm{r}>\rho_\mrm{rb}$
[white area in \fig{fig:phase-diagr}B].

When only the low risk fungicide (fungicide B) is applied, we set
$r_\mrm{B} = 1$ and, hence $\gamma_\mrm{s}=\gamma_\mrm{r}=1$ in the inequality (\ref{eq:brn-ineq-gen-app}) and obtain
\begin{equation}\label{eq:brn-ineq-fungica}
\frac{ C}{ C + C_{50}}   < \rho_\mrm{r}/k_\mrm{k} + \frac{ C}{ C +
    C_{50}} (1 - \rho_\mrm{r}),
\end{equation}
This inequality holds and the sensitive strain dominates for all positive values of $\rho_\mrm{r}$ and
$C$ at which $R_\mrm{0s}>1$.

Consider the case when the two fungicides A and B are applied
together at an arbitrary mixing ratio $r_\mrm{B}$, assuming no
pharmacological interaction ($u=0$) and perfect resistance of the
resistant strain to the fungicide A ($\alpha=0$).
In this case, $\gamma_\mrm{s} = 1$ and $\gamma_\mrm{r} =
r_\mrm{B}$. Substituting these values in \eq{eq:rhorb-gen-app},
\eq{eq:c12-gen-app} and \eq{eq:d-gen-app} gives the same expressions
as in \eq{eq:rhorb-fungica}, \eq{eq:c12-fungica} and
\eq{eq:d-fungica}, but with $\alpha$ substituted by $r_\mrm{B}$.

\subsection{Expressions for the treatment benefit}
\label{sec:app-trben}

The treatment benefit is defined as the ratio between the amount of
healthy hosts $H(t)$ when both the disease and treatment are
present and the amount of healthy hosts at no disease $B(t) =
H(t)/H_\mrm{nd}(t)$ (see \sec{sec:trben}). 

In order to obtain analytical expressions for $B(t)$, we consider one growing season and assume the host-pathogen
equilibrium is reached during the season.
 This corresponds to the time-dependent solution of
 Eqs.\,(\ref{eq:1host2fung-gen-1})-(\ref{eq:1host2fung-gen-3})
 reaching its stable fixed point (or steady state).  Fixed points of
 Eqs.\,(\ref{eq:1host2fung-gen-1})-(\ref{eq:1host2fung-gen-3}) can be
 found by equating the right-hand sides of all equations to zero and
 solving the resulting algebraic equations with respect to $H(t)$,
 $I_\mrm{s}(t)$ and $I_\mrm{r}(t)$. Biologically this occurs when the
 first positive term in \eq{eq:1host2fung-gen-1} corresponding to
 growth of healthy hosts, is compensated by the second, negative term
 that corresponds to the decrease in healthy hosts due to
 infection. In other words, equilibrium occurs when the rate of
 emergence of new healthy tissue as a result of plant growth is
 exactly offset by the rate of its decrease due to infection. The
 right-hand side of \eq{eq:1host2fung-gen-2} goes to zero, when the
 rate of increase in $I_\mrm{s}$ due to new infections is compensated
 by the loss of the infectious tissue due to the completion of the
 infectious period (similar reasoning applies
 for \eq{eq:1host2fung-gen-3}).

Then, the treatment benefit is given by
\begin{equation}\label{trben-gen}
B(t \to \infty) = B^* = \frac{H^*}{K},
\end{equation}
where $H^*$ is the equilibrium amount of healthy hosts and $K$ is the
host carrying capacity, and assume full resistance ($\alpha=0$).

When only the fungicide B is applied at a dose $C$
[\fig{fig:trben-vs-conc-cost-col}A], the basic reproductive number of
the sensitive pathogen strain always exceeds the one for the resistant
strain $R_\mrm{0s}>R_\mrm{0r}$. Therefore, the resistant mutants are
eliminated in the long run and the amount of the healthy host tissue is equal
to $H^* = \mu/ (b \left[ 1 - \veps(C) \right] )$, where $\veps(C)$ is given
by \eq{eq:eps-1fungic}. Then, according to \eq{trben-gen}, the
treatment benefit is
\begin{align}\label{eq:trben-fungicb}
B^*(C)=  \frac{ \mu } {b  \left[ 1 - \veps(C) \right] K}.
\end{align}
It grows with the fungicide dose and saturates, since the
function $\veps(C)$ saturates.

Application of the fungicide A alone at a dose $C$ may favor
either resistant or sensitive pathogen strain depending on the fitness
cost of resistance $\rho_\mrm{r}$ and the fungicide dose $C$
[see \fig{fig:phase-diagr}B]. The treatment benefit in this case is
\begin{align}\label{eq:trben-fungica}
B^*(C, \rho_\mrm{r})= 
\begin{cases} 
\frac{ \mu } { b \left[ 1-\veps(C) \right] K }, & \mrm{for} \:
(\rho_\mrm{r} <k_k\: \mrm{and} \:  C<C_\mrm{b} ) \\ & \mrm{or}
\: (\rho_\mrm{r} > k_k \: \mrm{and} \: \forall C),\\
\frac{\mu} {  b ( 1-\rho_\mrm{r}) K }, &  \mrm{for} \:
\rho_\mrm{r} < k_k \: \mrm{and} \: C>C_\mrm{b},
\end{cases}
\end{align}
where the $C_\mrm{b}$ is given by
\eq{eq:cb-fungica-perfres}.

Now, consider application of both fungicides in a mixture at equal
concentrations ($r_\mrm{B} = 1/2$), assuming no interaction between
fungicides ($u=0$). In this case, again either resistant or sensitive
pathogen strain will dominate the population depending on the fitness
cost $\rho_\mrm{r}$ and the total fungicide dose $C$ [see
\fig{fig:phase-diagr}C]. The treatment benefit now has the following
expression
\begin{align}\label{eq:trben-equalmix}
B^*(C, \rho_\mrm{r})= 
\begin{cases} 
   \frac{\mu} { b \left[1 - \veps(C) \right] K}, & \mrm{for} \: (\rho_\mrm{r}
  <\rho_\mrm{rb} \: \mrm{and} \: (C<C_\mrm{b1} \: \text{or}\:
  C>C_\mrm{b2})) \\ & \mrm{or}
  \: (\rho_\mrm{r} > \rho_\mrm{rb} \: \mrm{and} \: \forall C),\\
  \frac{\mu } { b \left[1 - \veps(C/2) \right]  (1 - \rho_r) K }, & \mrm{for} \:
  \rho_\mrm{r} <\rho_\mrm{rb} \: \mrm{and} \:
  C_\mrm{b1}<C<C_\mrm{b2},
\end{cases}
\end{align}
where $\rho_\mrm{rb}$, $C_\mrm{b1}$ and $C_\mrm{b2}$ and are given by
\eq{eq:rhorb-gen-app} and \eq{eq:c12-gen-app} at $\gamma_\mrm{s} =1$,
$\gamma_\mrm{r} = 1/2$. 
%
%
The treatment benefit
$B^*(C, \rho_\mrm{r})$ is shown as a function of the fungicide
dose $C$ and the fitness cost of resistance $\rho_\mrm{r}$ in
\fig{fig:trben-vs-conc-cost-col} for the three cases discussed above,
according to Eqs.\,(\ref{eq:trben-fungicb})-(\ref{eq:trben-equalmix}).

\begin{figure}[!ht]
  \centerline{\includegraphics[width=0.6\textwidth]{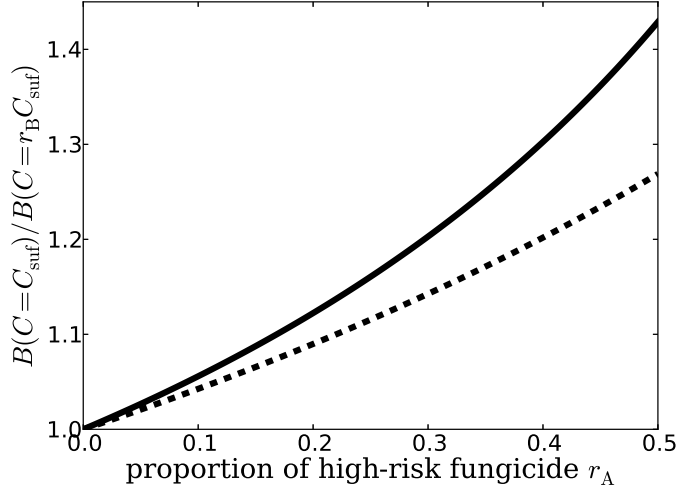}}
  \caption{
    Extra benefit of adding the high-risk fungicide to the mixture
    plotted according to \eq{eq:trben-rat} and \eq{eq:csuf} versus the
    proportion of the high-risk fungicide $r_\mrm{A}$, provided the
    sensitive pathogen strain is favored by selection. The basic
    reproductive number of the sensitive strain in the absence of
    fungicides $R_0 = b K / \mu$ had the value $R_0=2$ (dashed curve)
    and $R_0 =4$ (solid curve). Other parameters: $C_{50}=1$,
   $B_\mrm{suf}=0.9$, maximum fungicide effect on $b$ is
   $k_{\mrm{k}b}=0.9$ and the equivalent maximum fungicide effect on
   $\mu$ is $k_{\mrm{k}\mu}=k_{\mrm{k}b}/(1 - k_{\mrm{k}b}) = 9$ (see
   text for explanation).}
\label{fig:benrat-vs-ra}
\end{figure}

When a mixture of a high-risk and a low-risk fungicide is used with
the total dose $C$ and the proportion of the low-risk fungicide
$r_\mrm{B}$, a relevant question arises: ``How much additional control
does the addition of the high-risk fungicide provide?''.
In order to quantify the degree of additional control due the
high-risk fungicide component of the mixture, we first set the
sufficient level of control that needs to be achieved in terms of the
treatment benefit $B^*=B_\mrm{suf}$ (for example, we can set
$B_\mrm{suf}=0.9$). Then, we determine the dose $C_\mrm{suf}$ of the
fungicide mixture at which this level of control is achieved. We
assume that the proportion of the two fungicides in the mixture was
chosen such that $r_\mrm{B}>r_\mrm{Bc}$, i.\,e. the sensitive pathogen
strain is favored by selection. Hence, the the treatment benefit is
given by $B^*=\mu/(b K)$ [the upper expression in
\eq{eq:trben-equalmix}]. We set $B^*=B_\mrm{suf}$ and substitute the
dependence of the pathogen infectious period or the transmission rate
on the fungicide dose according to $\mu \to \mu ( 1 + \veps(C))$ or
$b \to b ( 1 - \veps(C))$. Then, we obtain
\begin{align}\label{eq:csuf}
  C_\mrm{suf} =
\begin{cases} 
   C_{50} \frac {B_\mrm{suf} R_0 - 1} {1 + k_\mrm{k} - B_\mrm{suf} R_0},
   & \mrm{if} \: \mrm{fungicides} \: \mrm{affect} \: \mu\\ 
   C_{50} \frac {B_\mrm{suf} R_0 - 1} {1 - B_\mrm{suf} R_0 (1 - k_\mrm{k}) }, &  \mrm{if} \: \mrm{fungicides} \: \mrm{affect} \: b.
\end{cases}
\end{align}
The ratio
%
\begin{align}\label{eq:trben-rat}
\frac {B^*(C=C_\mrm{suf})} {B^*(C=r_\mrm{B} C_\mrm{suf})} = 
\begin{cases} 
   \frac {1 + \veps(C_\mrm{suf})} {1 + \veps(r_\mrm{B} C_\mrm{suf})},
   & \mrm{if} \: \mrm{fungicides} \: \mrm{affect} \: \mu\\ 
   \frac {1 - \veps(r_\mrm{B} C_\mrm{suf}) } {1 - \veps(C_\mrm{suf}) }, &  \mrm{if} \: \mrm{fungicides} \: \mrm{affect} \: b
\end{cases}
\end{align}
characterizes the extra benefit due to addition of the high-risk
fungicide, since $B^*(C=C_\mrm{suf})$ is the treatment benefit when
both high-risk and low-risk fungicides are present and
$B^*(C=r_\mrm{B} C_\mrm{suf})$ is the treatment benefit when the
high-risk fungicide is absent (here, $r_\mrm{B}$ is the proportion of
the low-risk fungicide in the mixture). The ratio $B^*(C=C_\mrm{suf})
/ B^*(C=r_\mrm{B} C_\mrm{suf})$ is shown in \fig{fig:benrat-vs-ra} as
a function of the proportion of the high-risk fungicide in the mixture
$r_\mrm{A} = 1 - r_\mrm{B}$. One sees from \fig{fig:benrat-vs-ra} that
the more high-risk fungicide is used in the mixture, the larger is the
extra benefit from its application (provided that
$r_\mrm{B}>r_\mrm{Bc}$, when the sensitive pathogen strain is favored
by selection). However, the largest $r_\mrm{A}$ which still does not favor the
resistant strain is determined by the value of the fitness cost of
fungicide resistance (see \sec{sec:optrat}).

Interestingly, the ratio $B^*(C=C_\mrm{suf}) / B^*(C=r_\mrm{B}
C_\mrm{suf})$ does not depend on where the fungicide acts, on the
infectious period $\mu^{-1}$ or the transmission rate $b$, as long as
the maximum fungicide effects in these to cases $k_{\mrm{k}b}$ and
$k_{\mrm{k}\mu}$ are related by $k_{\mrm{k}\mu} = k_{\mrm{k}b}/(1 -
k_{\mrm{k}b})$ such that the basic reproductive number is reduced by
the same amount when the maximum effect is achieved.


\subsection{Dynamics of the frequency of the resistant pathogen strain}
\label{sec:app-delsel}

If the fungicide resistance is not associated with a fitness cost,
then the resistant strain is favored by selection and eventually
dominates the population whenever the high risk fungicide is applied
alone or in a mixture with the low risk fungicide [\fig{fig:phase-diagr}B,C].
However, for a given value of the total fungicide dose $C$,
the selection for resistance slows down when applying the fungicide
mixture as compared to the treatment with the high risk fungicide
alone [as seen from time-dependent numerical solutions of the model
Eqs.\,(\ref{eq:1host2fung-gen-1})-(\ref{eq:1host2fung-gen-3})] in
agreement with the findings of \cite{hopa+11a}.

In order to understand this result we consider the dynamics of the
frequency of the resistant pathogen strain $p(t) =
I_\mrm{r}/(I_\mrm{r} + I_\mrm{s})$. The rate of its change is obtained
from Eqs.\,(\ref{eq:1host2fung-gen-1})-(\ref{eq:1host2fung-gen-3}) \cite{mile+89}
\begin{equation}\label{eq:resfreq-dyn}
\frac{d p}{d t} = s(t) p (1 - p),
\end{equation}
where 
\begin{equation}\label{eq:sel-coeff-app}
  s = b \left[ (1 - \veps_\mrm{r}(C, r_\mrm{B})) (1 - \rho_\mrm{r} ) -
    (1 -\veps_\mrm{s}(C, r_\mrm{B})) \right] H(t)
\end{equation}
is the selection coefficient [a similar expression was found in
\cite{gugi99}]. Here $\veps_\mrm{s}(C, r_\mrm{B}) = k_\mrm{k} C / ( C
+ C_{50}/ \gamma_\mrm{s})$, $\veps_\mrm{r}(C, r_\mrm{B}) = k_\mrm{k} C
/ ( C + C_{50}/ \gamma_\mrm{r})$ and $r_\mrm{B}$ is the proportion of
the fungicide B in the mixture. Here, $C=C_\mrm{A} + C_\mrm{B}$, where
the dose $C_\mrm{A}$ of the fungicide A and the dose $C_\mrm{B}$ of
the fungicide B may depend on time due to fungicide decay:
\begin{equation}\label{eq:cab-t}
C_\mrm{A} = C_\mrm{A0} \exp \left( -\nu_\mrm{A} t \right), \: C_\mrm{B} = C_\mrm{B0} \exp \left( -\nu_\mrm{B} t \right)
\end{equation}
where $C_\mrm{A0}$, $C_\mrm{B0}$ are the fungicide doses at the time
of application, $\nu_\mrm{A}$ and $\nu_\mrm{B}$ are the fungicide
decay rates.

The expression (\ref{eq:sel-coeff-app}) for the selection coefficient was
obtained under the assumption that the fungicide decreases the
transmission rate $b$. In the case when the fungicide decreases the
infectious period $\mu^{-1}$, the selection coefficient does not depend on the
amount of healthy hosts $H(t)$.

Variables in \eq{eq:resfreq-dyn} can be separated and a
closed-form solution is found
\begin{equation}\label{eq:resfreq-dyn-sol}
\int \frac{d p}{p (1 - p)} = \int_0^{t_\mrm{m}} s( t ) dt.
\end{equation}

One can see from \eq{eq:resfreq-dyn-sol} that the overall selection
over the time $t_\mrm{m}$ is determined by the integral of the
selection coefficient $s(t)$ over time $\int_0^{t_\mrm{m}} s(t)
dt$. We are interested in the overall selection that occurs during the
time $t_\mrm{m}$ which is longer than the time scale of change in the
fungicide dose. In this case, an equivalent, constant over time
fungicide dose can be determined, which gives rise to the same value
of the integral $\int_0^T s(t) dt$. This effective fungicide dose
would take into account the time-dependent effect of the amount of the
host tissue on the strength of selection.

Assuming a zero fitness cost ($\rho_\mrm{r}=0$), no pharmacological
interaction ($u=0$) and full resistance ($\alpha=0$), the selection
coefficient can be written as 
\begin{equation}\label{eq:sel-coeff1}
s(t) = b \left[ \veps(C) - \veps(r_\mrm{B} C) \right] H(t).
\end{equation}
Assuming that $H(t)$ is a slowly varying function compared to the time
scale of selection, the solution of \eq{eq:resfreq-dyn} reads:
\begin{equation}\label{eq:resfreq-logist}
p(t) = \frac{p_0 \exp [s(t) t]}{ 1 + p_0 \left( \exp [s(t) t] - 1 \right)},
\end{equation}
where $p_0 = p(t=0)$. At $s>0$, the function $p(t)$ grows monotonically
and tends to one at large times. The rate, at which it grows is
determined by the magnitude of the selection coefficient $s$.

One can see from \eq{eq:sel-coeff1} that when the high risk fungicide is
applied alone ($r_\mrm{B} = 0$), the selection coefficient is larger
than when it is mixed with a low risk fungicide ($0<r_\mrm{B}<1$) at the same
total fungicide dose $C$. Hence, $s(r_\mrm{B}=0,C) >
s(r_\mrm{B}>0,C)$. This is because the function $\veps(r_\mrm{B}
C)$ has positive values for any
$r_\mrm{B}>0$. Thus, the selection for the resistant strain (against the
sensitive strain) is delayed when a mixture of high risk and low risk
fungicides is applied compared to treatment with the high risk
fungicide alone. A careful consideration of \eq{eq:resfreq-dyn-sol}
reveals that this conclusion holds also when $H(t)$ does not vary
slowly over the time scale of selection. Lower fungicide dose will
decrease the selection coefficient under the integral on the right-hand side of
\eq{eq:resfreq-dyn-sol}. Hence, in order to achieve a given large value
of the frequency of resistance $p$, one would need to integrate over a
longer time $t_\mrm{m}$.

\subsection{Generalization of the model: effect of the fungicide and fitness cost of resistance
  on the pathogen}
\label{sec:gener}


So far we assumed that both the resistance cost and fungicides affect
the transmission rate $b$. We performed the same analysis for the
three remaining cases possible in the model: When (i) both resistance
cost and the fungicide affect the pathogen death rate according to
$\mu \to \mu (1 + \rho_\mrm{r} + \veps_\mrm{r}(C, r_\mrm{B}))$ for the
resistant strain and $\mu \to \mu (1 + \veps_\mrm{s}(C, r_\mrm{B}))$
for the sensitive strain; (ii) the resistance cost affects the
transmission rate $b \to b (1 - \rho_\mrm{r})$ of the resistant strain
and the fungicides affect the pathogen death rate $\mu \to\mu (1 +
\veps_\mrm{s,r}(C, r_\mrm{B}))$ ; (iii) resistance cost affects the
death rate of the resistant pathogen strain $\mu \to \mu (1 +
\rho_\mrm{r})$, while the fungicide affects the infection rate of both
resistant and sensitive strains $b \to b (1 - \veps_\mrm{r,s}(C, r_\mrm{B}))$.
We have found that although the mathematical expressions for the
results have a different form in these cases and there is a slight
quantitative difference, all the conclusions remain the same and do
not depend on whether the fungicide and the resistance cost manifest
in the infection rate $b$ or in the pathogen death rate $\mu$.

Moreover, we have done the same analysis using a fungicide
dose-response function different from \eq{eq:eps-1fungic}, namely
using the function $\veps(C) = \veps_\mrm{m} ( 1 - \exp
\left[ - \beta C \right])$. If the two fungicides have the same values
of $\veps_\mrm{m}$ and $\beta$ and are applied at doses $C_\mrm{A}$
and $C_\mrm{B}$, then according to Loewe's additivity, their combined
action has the form $\veps(C_\mrm{A}, C_\mrm{B}) = \veps_\mrm{m} ( 1 -
\exp \left[ - \beta (C_\mrm{A} + C_\mrm{B}) \right])$. We found again
that all the conclusions remain the same in this case.

This generalization applies to determination of the direction of
selection (the sign of the selection coefficient in
\eq{eq:resfreq-dyn}) and to the outcomes for the treatment benefit at
equilibrium obtained in \sec{sec:trben}. However, the time-dependent
solutions of
Eqs.\,(\ref{eq:1host2fung-gen-1})-(\ref{eq:1host2fung-gen-3}) may
behave differently depending on how the fungicide and the fitness cost
affect the pathogen life cycle and the form of the fungicide
dose-response function. This is an interesting topic for further
investigations, but lies beyond the scope of this study.

\subsection{Fungicide mixture versus alternation}
\label{sec:app-mix-vs-altern}

It was previously discussed \cite{sh06} that in the presence of a
fitness cost the alternation of fungicides can be effective, but we
have shown here that fungicide mixtures can also be effective in this
case. When using an alternation strategy, the period of selection
during which the resistant strain is favored in the presence of the
high risk fungicide is followed by a period during which selection
favors the sensitive strain in the absence of this fungicide. The
latter period is typically much longer because the selection pressure
induced by the high risk fungicide is much larger than that induced by
the fitness cost of resistance. Hence, one needs to wait for quite a
long time before the resistant strain disappears and the high risk
fungicide can be used again. Moreover, there are times during which
the frequency of the resistant strain becomes large (at the end of the
period of the application of the high risk fungicide), which increases
the risk that resistance will spread to other regions. Both of these
disadvantages are avoided by using a mixture where the proportion of
the low risk fungicide is above a critical value determined here
(\fig{fig:opt-fung-rat-comb}). In this case there is no need to delay
the application of the high risk fungicide and the frequency of the
resistant strain does not rise above the mutation- or
migration-selection equilibrium because the mixture does not induce
selection for resistance.

\subsection{The risk of double resistance}
\label{sec:dr-risk}

Although we do not consider the possibility of double resistance in
our model, by applying an optimal proportion of fungicides in the
mixture as suggested here, one would prevent selection for resistance
to the high risk fungicide. 
Consequently, the risk of development of double resistance would be
reduced.
For both sexually and asexually reproducing pathogens, there are three
pathways for generating double resistance: (i) A-resistant
mutants are produced first and then a proportion of them acquires also
B-resistance by spontaneous mutation (ii) B-resistant mutants are
generated first and subsequently acquire A-resistance and (iii) double
resistance is generated directly from the wild-type. In this case, by
preventing selection for A-resistance, one removes only the pathway (i)
to double resistance. 
If a pathogen is able to reproduce sexually, then a much more
likely scenario for the double resistance to emerge is through
recombination. For the recombination to occur, both singly resistant
strains (A-resistant and B-resistant) would need to be present in the
population at significant frequencies. Hence, preventing selection for
A-resistance would diminish the probability of the emergence of double
resistance by recombination.
Thus, our findings would also help to significantly reduce the risk of
development of double resistance, especially in sexually reproducing
pathogens.



\end{document}